\documentclass[useAMS,usegraphicx]{mn2e}
\usepackage{rotate}
\usepackage{rotating}
\usepackage{times}
\usepackage{cite}             
\usepackage{comment}  
\usepackage{graphicx}    
\newif\ifAMStwofonts
\AMStwofontstrue

\def\he0436{{HE~0436--4717}}

\def\kdblur2{\emph{kdblur2}}

\def\kdblur{\emph{kdblur2}}

\def\alphaox{$\alpha_{\rm{ox}}$}                                                        
\def\indexin{$\alpha_{\rm{in}}$}       
\def\indexout{$\alpha_{\rm{out}}$}     
\def\Rin{{$R_{\rm{in}}$}}
\def\Rout{{$R_{\rm{out}}$}}
\def\Rbr{{$R_{\rm{br}}$}}
\def\Rcor{{$R_{\rm{cor}}$}}        
\def\Fvar{{$F_{\rm{var}}$}}

\def\Msun{M$_{\odot}$}
\def\Mbh{\thinspace \emph{M$_{\rm{BH}}$}}

\def\chandra{{\it Chandra }}

\def\rosat{{\it ROSAT }}
\def\euve{{\it EUVE }}
\def\asca{{\it ASCA }}

\def\swift{{\it Swift }}
\def\xmm{{\it XMM-Newton }}
\def\astroh{{\it Astro-H }}            

\def\et{{et al.\ }}
\def\rg{{\thinspace $R_{\rm g}$}}

\def\chisq{{$\chi^2$}}                     
\def\chidof{{$\chi^2_\nu/{\rm dof}$}}
\def\redchi{{$\chi^2_\nu$}}
\def\delchi{{$\Delta\chi^2$}}
\def\feka{{Fe~K$\alpha$}}

\def\ovii{{O~\textsc{vii}}}
\def\oviii{{O~\textsc{viii}}}

\def\nh{{$N_{\rm H}$}}   
\def\hb{{H$\beta$}}         

\def\Ledd{{$L_{\rm{Edd}}$}}              
\def\Rg{{$\rm\thinspace R_{g}$}}     
\def\deg{$^{\circ}$}

\def\A{{\rm\thinspace \AA}}
\def\cm{{\rm\thinspace cm}}
\def\erg{{\rm\thinspace erg}}
\def\eV{{\rm\thinspace eV}}

\def\ga{{\rm\thinspace gauss}}

\def\keV{{\rm\thinspace keV}}
\def\km{{\rm\thinspace km}}
\def\kpc{{\rm\thinspace kpc}}

\def\Mpc{{\rm\thinspace Mpc}}
\def\Msun{{\hbox{$\rm\thinspace M_{\odot}$}}} 

\def\s{{\rm\thinspace s}}
\def\arcsec{{\rm\thinspace arcsec}}   
\def\ks{{\rm\thinspace ks}}
\def\ps{{\rm\thinspace s^{-1}}}

\def\cts{{\rm\thinspace count}}

\def\cps{\hbox{$\cts\s^{-1}\,$}}
\def\cmps{\hbox{$\cm\s^{-1}\,$}}

\def\ergpcmsps{\hbox{$\erg\cm^{-2}\s^{-1}\,$}}                         
\def\ergcmps{\hbox{$\erg\cm\s^{-1}\,$}}                                     
\def\ergpscmps{\hbox{$\erg\cm^{-2}\s^{-1}\,$}}
\def\ergps{\hbox{$\erg\s^{-1}\,$}}

\def\plnorm{\hbox{$\rm{ph}\keV^{-1}\cm^{-2}\s^{-1}\,$}}      

\def\kmps{\hbox{$\km\ps\,$}}
\def\kmpspMpc{\hbox{$\kmps\Mpc^{-1}\,$}}

\def\pscm{\hbox{$\cm^{-2}\,$}}

\title[Is HE~0436-4717 Anemic?]{Is HE~0436-4717 Anemic? A deep look at a bare Seyfert 1 galaxy}

\author[K. Bonson, L. C. Gallo, R. Vasudevan]{K. Bonson$^1$, L. C. Gallo$^1$ and R. Vasudevan$^2$ 
               \\ 
$^{1}$ Department of Astronomy and Physics, Saint Mary's University, 923 Robie Street, Halifax, NS, B3H 3C3, Canada \\
$^{2}$ Institute of Astronomy, Madingley Road, Cambridge, CB3 0HA \\
}
\date{Accepted. Received. }
\pagerange{\pageref{firstpage}--\pageref{lastpage}}
\pubyear{2014}
\begin{document}
\maketitle
\label{firstpage}

\begin{abstract}
A multi-epoch, multi-instrument analysis of the Seyfert~1 galaxy \he0436\ is conducted using optical to X-ray data from \xmm and \swift (including the BAT).
Fitting of the UV-to-X-ray spectral energy distribution shows little evidence of extinction and the X-ray spectral analysis does not confirm previous reports of deep absorption edges from \oviii.  \he0436\ is a ``bare'' Seyfert with negligible line-of-sight absorption making it ideal to study the central X-ray emitting region. Three scenarios were considered to describe the X-ray data: partial covering absorption, blurred reflection, and soft Comptonization. All three interpretations describe the 0.5--10.0\keV\ spectra well. Extrapolating the models to $100\keV$ results in poorer fits for the the partial covering model. When also considering the rapid variability during one of the \xmm observations, the blurred reflection model appears to describe all the observations in the most self-consistent manner. If adopted, the blurred reflection model requires a very low iron abundance in \he0436. We consider the possibilities that this is an artifact of the fitting process, but it appears possible that it is intrinsic to the object. 

\end{abstract}



\section{Introduction}
\label{sect:intro}

Seyfert 1 galaxies are a subclass of active galactic nuclei (AGN) that provide a preferred view of the central engine.  Seyfert 1 galaxies that appear to have little to no intrinsic absorption, so-called ``bare" Seyfert 1s, provide an especially unobscured line of sight to the innermost regions close to the supermassive black hole itself. Such AGN have been noted before (e.g. Vaughan et al. 2004, Emmanoulopoulos et al. 2011) and recently have become the subject of larger sample studies (Patrick et al. 2011, Jin et al. 2012, Walton et al. 2013). 

\he0436 appears at first to be a rather typical Seyfert 1 galaxy at \emph{z} = 0.053 first observed by the Einstein Slew Survey in 1992 (ESS, Elvis et al. 1992). It has been included in numerous surveys and catalogues since the ESS including: the \rosat All-Sky Survey (Grupe et al. 1998), the Hamburg/ESO Survey for bright QSOs (Wisotzki et al. 2000), and the Palermo Swift-BAT Hard X-ray Catalogue (Cusumano et al. 2010). Despite its frequent inclusion in such wide-field studies, \he0436 has been the subject of deep X-ray analysis only once (Wang \et 1998). 

\begin{table*}
\caption[HE 0436-4717 Data Log.]{HE 0436-4717 Data Log.}
\centering    
\scalebox{1.0}{                                      
		\begin{tabular}{ccccccc}
		\hline
		(1) & (2) & (3) & (4) & (5) & (6) & (7) \\
		\bf{Telescope} & \bf{ObsID}& \bf{Instrument} & \bf{Filter} & \bf{Start Date} & \bf{Duration} \rm{(s)} & \bf{GTI} \rm{(s)} \\\hline
		\xmm & 0112320201 & pn & thin & 09/10/2002 & 68177 & 67210 \\
		&& MOS 1 & thin & & 69167 & 68270 \\
		&& MOS 2 & thin &  & 69167 & 68380 \\
		&& OM & UVW1 &  & 34996$^{*}$ & N/A \\
		&&& V &  & 25000$^{*}$ & N/A \\
		& 0603460101 & MOS 1 & thin & 12/15/2009 & 129141 & 126900 \\
		&& MOS 2 & thin & & 129146 & 123600 \\
		\\
		\swift & 00035763001 & XRT & none & 12/07/2007 & 10090 & 9732 \\
		&& UVOT & V && 838 & 838 \\
		&&& B && 840 & 840 \\
		&&& U && 840 & 840 \\
		&&& UVW1 && 1681 & 1681 \\
		&&& UVM2 && 2286 & 2286 \\
		&&& UVW2 && 3369 & 3369 \\
		& 70-mo survey & BAT & none & 09/21/2010 & N/A & N/A \\
		\\
		\hline
		 \multicolumn{6}{c}{*Individual filters exposed for a range of 1998 - 5001\thinspace{s} each with a sum of 10 exposures for UVW1 and 5 exposures for V.}
		\end{tabular}
}
	\label{MyObs}
\end{table*}

Halpern \& Marshall (1996) monitored the object in the extreme UV with the \euve satellite for 20 continuous days and were able to produce a long light curve. The timing analysis of the light curve revealed possible flaring with a period of about 0.9 days. In addition, the authors also analyzed spectra from \emph{EUVE}'s short-wavelength spectrometer and found no emission line features in the 70 -- 110\A\ range. The column density listed in that work is comparable to Galactic extinction alone, indicating no intrinsic extinction in the object was observed.

A detailed X-ray analysis of this object was done by Wang et al. (1998) using data from \asca and \rosat observations. The authors found the 0.4 -- 10.0\keV\ spectrum was well modelled by a single power law with slope of $\Gamma$ = 2.65$\pm$0.20 or by a sum of a $\Gamma$ = 2.15$\pm$0.04 power law and a weak black body of temperature 29.0$\pm2.0$\eV. The authors describe an \feka\ emission feature with an equivalent width of around 430\eV\ and two absorption edges: one at \emph{E} $\sim$ 0.28\keV\ with an optical depth of $\tau$ $\sim$ 3.46 measured from the \rosat spectrum, and one at \emph{E} $\sim$ 0.83\keV\ with $\tau$ $\sim$ 0.26 from the \asca spectra -- the latter of which was attributed to a combination of \ovii\ and \oviii\ edges. A comparison between \asca observations 4 months apart saw a 47\% flux change, however all other spectral parameters remained constant.

In this study, an analysis of \he0436 multi-epoch data from \xmm and \swift observatories spanning from 2002 -- 2013 is presented. \chandra also serendipitously observed this object between 2000 -- 2006, however these data were determined to be piled up and were not used in the analysis (see Section \ref{chandra}). The study of \xmm and \swift data presented here is the deepest investigation of this object to date using the most current instruments and analysis tools.

Following this introduction, Section \ref{data} describes the data processing for each observation. An initial X-ray spectral analysis of the \xmm EPIC data is described in Section \ref{firstmodeling}, including a search for absorption features seen previously in this object. The spectral analysis was extended into the UV band in Section \ref{UVSED}, which includes spectral energy distribution (SED) modelling using \swift data (UVOT, XRT, and BAT). Section \ref{XraySpec} describes an in-depth X-ray analysis of the \xmm MOS spectra considering absorption (Sec. \ref{absorption}), reflection (Sec. \ref{sec:meanref}), and soft Comptonization (Sec. \ref{SoftCompt}) scenarios. Section \ref{Timing} details the timing analysis of the \xmm light curves and specifically examines hardness ratios (Sec. \ref{HRs}), fractional variability (Sec. \ref{Fvar}), and flux-resolved spectra (Sec. \ref{FRS}). A discussion of the findings (Sec. \ref{Discussion}) and final concluding remarks (Sec. \ref{Conclusions}) complete this work.


\section{Observations and data reduction}
\label{data}


\subsection{\xmm}
\label{XMM}
\xmm (Jansen \et 2001) observations of \he0436 were serendipitous during the targeted observation of the pulsar PSR J0437.4-4711 about 4.2 arcmin away. The first observation (hereafter XMM1) was during revolution 0519 on the $9^{th}$ of October 2002 and spanned 70.5\ks. Two MOS detectors (Turner \et 2001) operated in full-frame mode with the thin filter. Due to the variable nature of the pulsar target, the pn camera (Str\"uder \et 2001) was set to timing mode also with a thin filter in place. The Optical Monitor (OM; Mason \et 2001) simultaneously collected data from \he0436 during this observation although the object was outside the field of view of the reflection grating spectrometers (den Herder \et 2001; RGS1 and RGS2). The OM data were limited to observations in the UVW1 ($\lambda_{\rm{max}}$ = 2675\A) and V ($\lambda_{\rm{max}}$ = 5235\A) filters. 

A second \xmm observation (hereafter XMM2) was during revolution 1835 on the $15^{th}$ of December 2009 and spanned 130\ks. The EPIC instruments operated in the same manner as in XMM1. During XMM2 only the MOS instruments observed \he0436 as the source fell outside of the pn field of view used for timing mode.

Data files from both epochs were processed to produce calibrated event lists using the \emph{XMM-} \emph{Newton} Science Analysis System (SAS) version 12.0.0. The data were examined for background flaring and pileup; only minor flaring was seen in the very beginning of XMM2 and those periods were ignored. No pileup was detected as the source was off-axis and the pn was operated in timing mode. For the MOS data, source photons were extracted from a circular region 35\arcsec\ in radius and centered on the object. The background photons were extracted from an area 50\arcsec\ in radius close to the object and then scaled appropriately. 

Source photons from the pn data were extracted from a rectangular region 2\deg\ in width centered on the object and the background photons from a larger rectangular region and also scaled. Single-quadruple events were selected for the MOS data while single and double events were selected for the pn; events next to a bad pixel or the CCD edge were omitted (i.e. data quality flag set to zero). The entire process was then repeated for XMM2. The resulting mean count rates were 3.07\cps\ for pn, 1.21\cps\ for MOS1, and 1.29\cps\ for MOS2 in the 0.3 -- 10.0\keV\ band.

Spectra from the MOS data were limited to the 0.5 -- 10.0\keV\ range based on cross-calibration uncertainties, while the pn spectra were limited to 0.6 -- 8.0\keV\ because of the less accurate calibration in timing mode\protect\footnote{See \xmm Calibration Technical Note 0018 (XMM-SOC-CAL-TN-0018) section 11.2.2 for details} and high background above 8\keV.


\subsection{\chandra}
\label{chandra}
The pulsar PSR J0437.4-4711 is a calibration source for the \chandra observatory. Of the eleven observations between 2000 -- 2006, \he0436 falls in the field of view during six: ObsIDs 742, 1850, 6155, 6156, 6157, and 7216. ObsIDs 742 and 1850 were observed with the HRC-I and -S detectors and thus only imaging data are available; these were ignored. The ACIS-S data were processed using the most recent version of CIAO at the time (v. 4.4.0) and preliminary spectra were extracted. After analyzing the spectra it was clear that the data were piled up. Several reprocessing methods were attempted, however the quality of the data were not high enough to justify any further analysis.

\subsection{\swift}
\label{Swift}
\he0436 has been observed by the \swift X-ray Telescope (XRT; Burrows \et 2005) twice in 2007 and most recently in June 2013 (hereafter Sw3). The first observation (hereafter Sw1) had the highest exposure time, followed by the second observation (hereafter Sw2). Processed Burst Alert Telescope (BAT) spectra and the diagonal response matrix are taken from the \swift\ BAT 70-Month Hard X-ray Survey (Baumgartner \et 2014). The BAT data are averaged spectra from integration of survey data collected over a 70-mo period, whereas the XRT observations are individual exposures of a few kiloseconds during pointed observations (see Table \ref{MyObs}).

The XRT observations were performed in photon counting mode.  X-ray data were reduced with the {\em xrtpipeline} version 0.12.1 and spectra were extracted with {\sc xselect}.  Optical and UV photometry was obtained with the UV-Optical Telescope (UVOT; Roming \et 2005).  Magnitudes and fluxes were based on the most recent UVOT calibration as described in Poole \et (2008) and Breeveld \et (2010).

A summary of the data used in this work is shown in Table \ref{MyObs}.


\section{Preliminary Analysis of XMM Data}
\label{firstmodeling}
All spectral model fitting was performed using the X-ray spectral fitting package XSPEC v. 12.8.1. Model parameters are reported in the rest frame of the AGN ($z=0.053$) and a cosmology of $H_{\rm{o}}$ = 70\kmpspMpc, $q_{\rm{o}}$ = 0, and $\Lambda_{\rm{o}}$ = 0.73 is assumed.
All models include a Galactic column density of $N_{\rm{H}}$ = 1.03$\times$$10^{20}$\thinspace$\rm{cm^{-2}}$ as determined from the LAB Survey\protect\footnote{http://heasarc.nasa.gov/cgi-bin/Tools/w3nh/w3nh.pl} (Kalberla \et 2005). Errors on model parameters correspond to a 90\% confidence level.

\begin{figure}
   \centering
   {\scalebox{0.3}{\includegraphics[angle=270,trim= 1cm 0cm 0.5cm 0.5cm, clip=true]{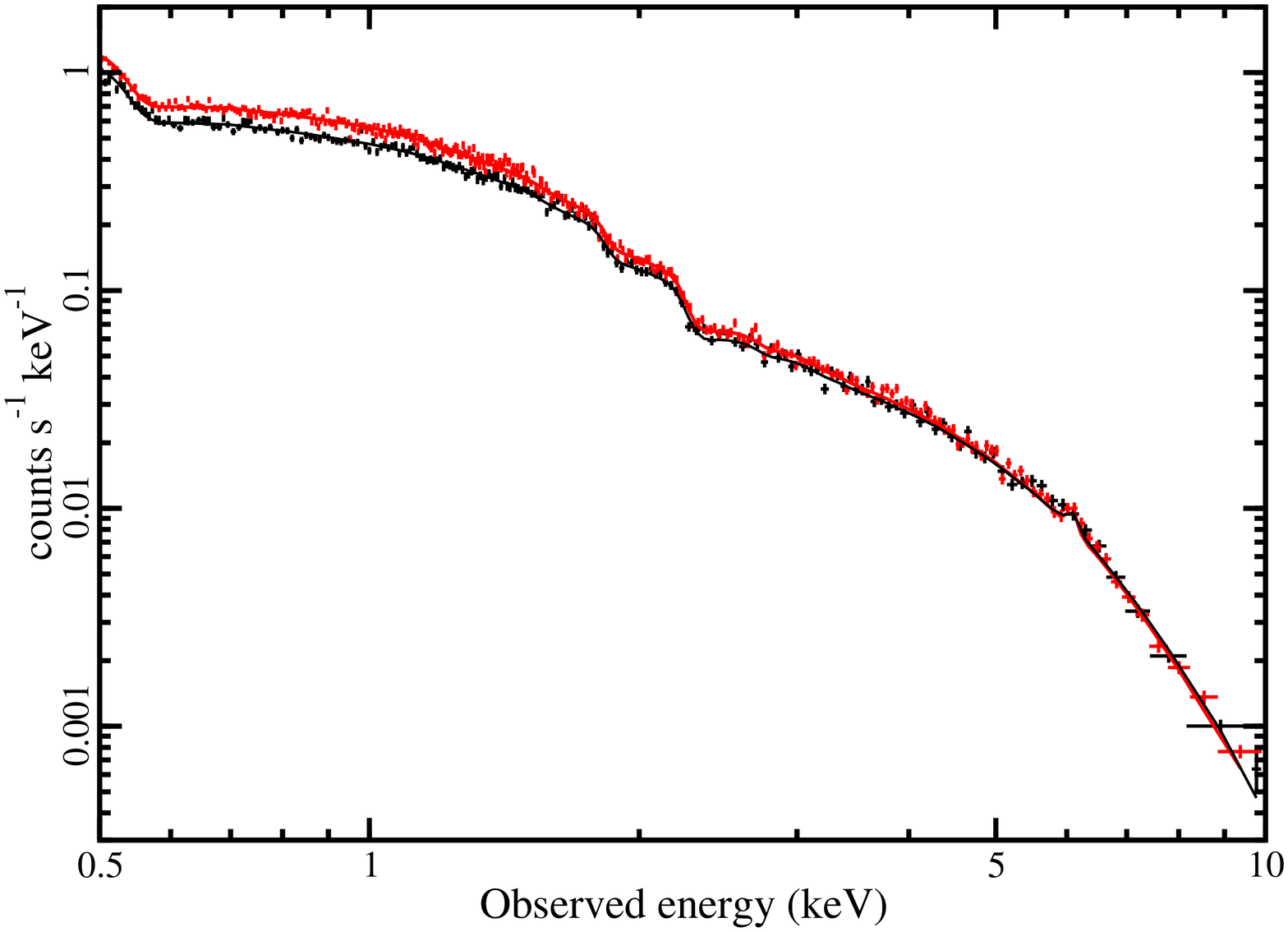}}} 
   {\scalebox{0.3}{\includegraphics[angle=270,trim= 1cm 0cm 0.5cm 0.5cm, clip=true]{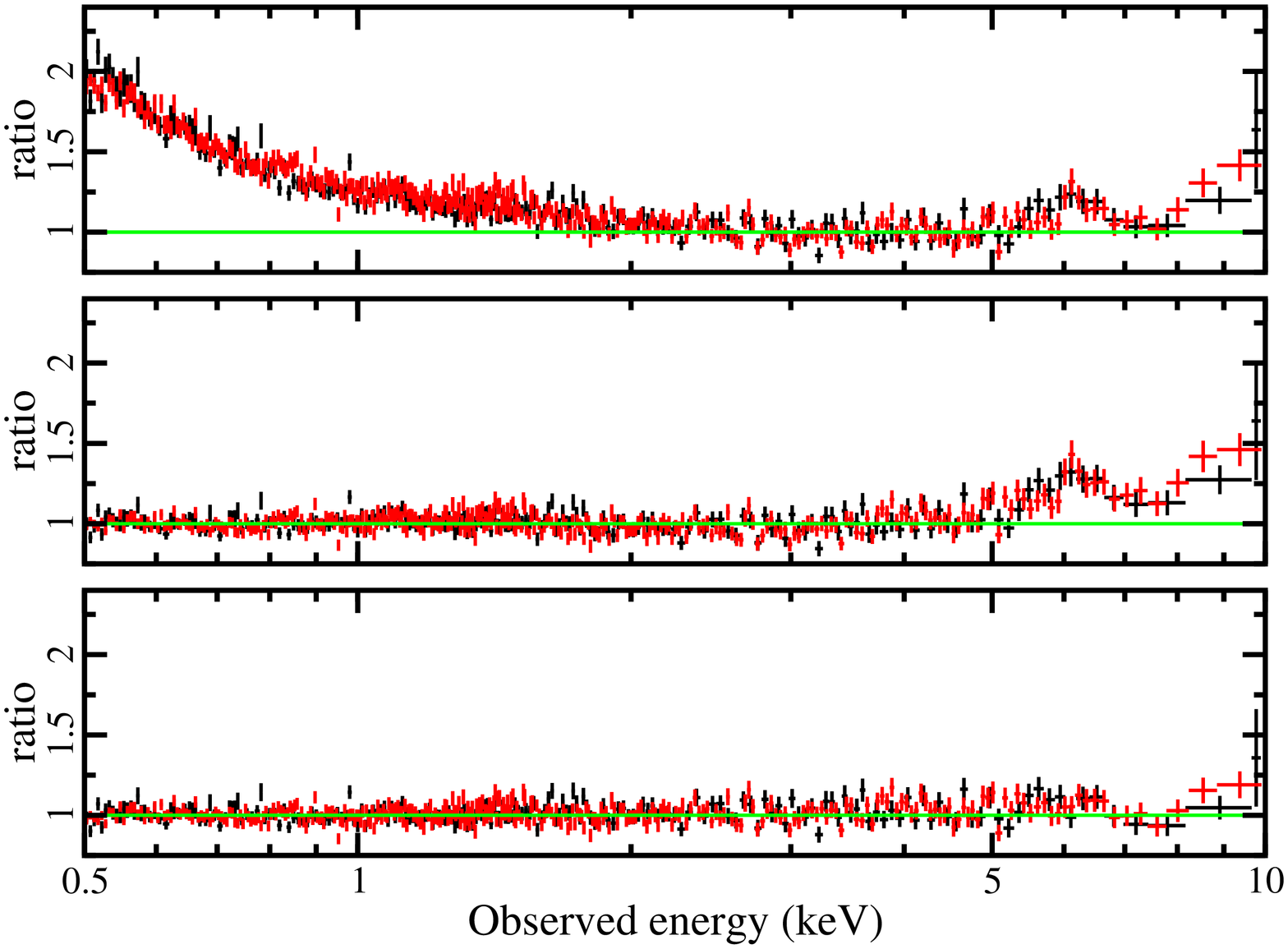}}} 
   \caption{Preliminary analysis of the \xmm merged MOS spectra. XMM1 (black) and XMM2 (red) data were fit with a power law plus black body and two Gaussians in the 0.5 -- 10\keV\* band (top panel). Extrapolating a power law fit to the $2-10\keV$ spectra to lower energies shows clearly a soft-excess below about 2\keV\* and and excess residuals around 6.4\keV\* (second from top). Adding a black body accounts for the soft-excess and residuals in the iron band (second panel from bottom). Including a Gaussian profile at about 6.4\keV\* with variable width reduces the residuals further (bottom panel).}
   \label{firstfit}
\end{figure}

\begin{figure}
   \centering
   {\scalebox{0.32}{\includegraphics[trim= 1cm 6cm 0cm 5.1cm, clip=true]{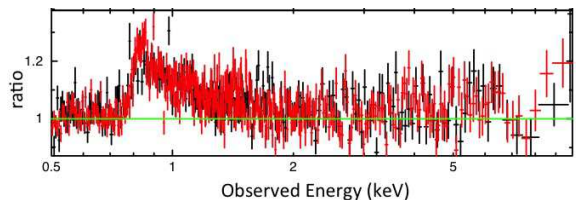}}} 
   \caption{An absorption edge with the same parameter values as those found in Wang et al. (1998) analysis of \asca spectra was added to the model and the residuals (data/model) are shown. These data clearly do not require the absorption edge and allowing the parameters free to vary returned an absorber with negligible optical depth. XMM1 is black, XMM2 is red.}
   \label{absedgetest}
\end{figure}

An initial inspection of the \xmm EPIC spectra showed consistency between the pn and MOS data for XMM1 and both MOS spectra for XMM2 within known calibration uncertainties. Since the multi-epoch data --- for which only MOS data are available --- will ultimately be compared, the pn data will not be discussed, although it was continually examined for consistency throughout the study. The MOS 1 and MOS 2 data at each epoch were merged into single spectrum using the FTOOLS program \textsc{addascaspec} in an effort to increase signal to noise and simplify model fitting.

The merged MOS spectra at each epoch were fit from 2 -- 10\keV\* with a power law and the model extrapolated to 0.5\keV.  The residuals clearly show a soft-excess, common in Seyfert 1 galaxies, and excess residuals around 6.4\keV\* (second from top panel, Fig.\thinspace \ref{firstfit}). The addition of a black body at low energies fit the soft-excess very well leaving only residuals at higher energies (second from bottom, Fig.~\ref{firstfit}). The addition of a narrow ($\sigma=1\eV$) Gaussian profile at $\sim6.4\keV$, common at both epochs, improved the fit further but still left significant residuals. Adding a second, broader Gaussian profile ($E\approx6\keV$ and $\sigma \approx3\keV$) improved the quality of fit (\delchi $= 14$ for 3 additional free parameters; Fig.~\ref{firstfit}, top panel) and residuals (Table~\ref{tab:firstfit}; Fig.~\ref{firstfit}, bottom panel). Reversing the order of line fitting --- adding a broad Gaussian line first --- does not influence the final fit as the fit statistic remained the same (i.e. two Gaussian profiles are still required to reconcile all iron residuals).

The described model provides a good fit to the multi-epoch spectra (\chidof\ $=1.07/2025$; Fig.~\ref{firstfit} top and bottom panels). Wang \et (1998) reported an absorption edge around \emph{E} $\sim$ 0.83\keV\ with $\tau$ $\sim$ 0.26 from their analysis of \asca spectra. Including this exact feature in the \xmm\ spectrum degraded the fit (Fig.\thinspace \ref{absedgetest}). Allowing the absorption energy and depth to vary did not improve the fit significantly (\delchi $= 11$ for 2 additional free parameters) and the measured edge energy ($E = 0.72\pm0.03$) and optical depth ($\tau= 0.06\pm 0.03$) were inconsistent with the \asca\ measurements.  Likewise, adding an absorber associated with the host galaxy did not enhance the fit and only an upper limit of \nh\thinspace $< 10^{19}$\pscm could be measured. The \xmm spectra here do not support significant additional absorption that was reported from the \asca analysis (Wang \et 1998).

The best-fit phenomological model presented in Table~\ref{tab:firstfit} does not reveal the physical processes at work in \he0436.  The broad Gaussian profile could be associated with blurred reflection or partial covering absorption. The strength of the soft-excess ($SE$), defined as the flux ratio between the black body and power law component between $0.4-3\keV$, could indicate whether the low-energy spectrum is sculpted by absorption or reflection (Vasudevan \et 2014), but the ratio measured in \he0436 ($SE\approx 0.1$) is consistent with either process. In the following sections we examine a physical description for the X-ray emission in \he0436.

\begin{table*}
\caption{The best phenomenological model fit to the merged \xmm MOS spectra. The model components and model parameters are listed in Columns 1, and 2, respectively. Columns 3 and 4 list the parameter values during the XMM1 epoch and XMM2 epoch, respectively. Fluxes are corrected for Galactic absorption and are reported in units of \ergpscmps. The soft-excess flux is measured between 0.4 -- 3.0\keV. Parameters linked across the epochs are denoted by dots and those without errors are frozen at the given values.}
\centering
\scalebox{1.0}{
	\begin{tabular}{ccccc}                
	\hline
	(1) & (2) & (3) & (4)  \\
 	Model Component &  Model Parameter  &  XMM1 & XMM2 \\
	\hline
 	Black body & $kT$ (\eV) & $77\pm 6$ & $79\pm 5$  \\
	&$F_{0.4-3 keV}$ & 0.68$\pm0.05\times10^{-12}$ & 0.72$\pm0.05\times10^{-12}$    \\
	\\
	Power law & $\Gamma$ & $2.14\pm 0.04$ & $2.23\pm0.03$ \\
	&$F_{0.4-3 keV}$ & 5.36$\pm0.05\times10^{-12}$ & 6.71$\pm0.05\times10^{-12}$    \\
	\\
	Narrow & $E$ (\keV) & $6.42\pm 0.06$ & \ldots \\
	Gaussian & $\sigma$ (\keV)  & $1$  & \ldots \\
	         & $EW$ (\eV)  & $46$  & $44$  \\
	         \\
	Broad  & $E$ (\keV) & $6.05^{+0.82}_{-3.36}$ & \ldots \\
	Gaussian & $\sigma$ (\keV)  & $3.6^{+3.9}_{-1.1}$  & \ldots \\
	         & $EW$ (\keV)  & $2.5$  & $2.3$  \\
	         \\
	Observed  & $F_{2-10 keV}$ & $4.30\times10^{-12}$ & $4.65\times10^{-12}$    \\ 
	Flux      & $F_{0.5-10 keV}$ & $8.22\times10^{-12}$ & $9.52\times10^{-12}$    \\
	\\
	Soft- & $F_{\rm{bb}}$ / $F_{\rm{pl}}$ & $0.13\pm 0.01$ & $0.11\pm0.01$ \\
	Excess & &  &     \\
	\\
         Fit Quality & \chidof & 1.07 / 2025    &  \\
         \hline
\label{tab:firstfit}
\end{tabular}
}
\end{table*}


\begin{table*}
\centering{   
\caption{The four models used to fit the UV-to-X-ray SED. Column 1 lists the model parameters and goodness-of-fit. Column 2 shows the parameter values for each model. Parameters that are fixed are listed without errors.
}} 
\scalebox{1.0}{       
		\begin{tabular}{ccccc}
		\hline
		(1) && \multicolumn{2}{c}{(2)} \\
		&& \multicolumn{2}{c}{\bf{SED Models}} & \\
		\bf{Parameters} & \bf{A} & \bf{B} & \bf{C} & \bf{D}  \\\hline
		 $L_{0.001-100\keV}$ & 6.39 &  15.17 & 6.47 & 4.99  \\
		 (\small{$\times10^{44}$\ergps}) &&&& \\
		 $E_{B-V,int}$ & 0 & 0.095$\pm$0.011 & 0.04 & 0  \\
		 $E_{B-V,Gal}$ & 0.009 & 0.009 & 0.009 & 0.009 \\
		 $T_{max}$ \small{(\eV)} & 5.26$\pm$0.01 & 6.85$\pm$0.21 & 3.11$\pm$0.25 & 2.35$\pm$0.14 \\
		\emph{K} \small{($\times10^{4}$\Msun$^2\kpc^{-2}$)} & 6.41 & 6.41 & 52.87$\pm$14.30 & 107.54$\pm$25.77 \\
		 power law norm. \small{($\times10^{-2}$)} & 1.36$\pm$0.06 & 0.63$\pm$0.06 & 6.22$\pm$1.42 & 14.16$\pm$2.59  \\
		 \small{(\plnorm)} &&&& \\
		 $L_{\rm{Bol}}$ / \Ledd & 0.08 & 0.20 & 0.09  & 0.07 \\
		  \redchi / d.o.f. & 3.50 / 30 & 1.06 / 29 & 1.03 / 29 & 1.02 / 29  \\
		  \hline
		\end{tabular}
}
\label{SEDmodels}
\end{table*}

\section{UV/X-ray Spectral Energy Distribution}
\label{UVSED}
X-ray emission provides valuable insight into the accretion process in the inner regions closest to the black hole, however UV radiation from the accretion disk dominates the bolometric emission. The optical/UV and X-ray bands are investigated together to capture as much of the total accretion emission as possible. One objective in such broadband analysis is to determine whether or not an AGN is potentially bare.

\subsection{The UV-to-X-ray spectral slope}
The slope of a hypothetical power law connecting the X-ray (2\keV) and UV (2500\A) spectral bands (\alphaox, Tananbaum et al. 1979) was estimated for \he0436 at both epochs using simultaneous \xmm MOS and OM data from the UVW1 filter. The UV luminosity at 2500\A\ is $\sim10^{29}$\ergps. The \alphaox\ values were comparable at both epochs with  \alphaox$_{\rm{,XMM1}}$ = -1.27 and \alphaox$_{\rm{,XMM2}}$ = -1.26. The higher X-ray flux in XMM2 predicts that the \alphaox\ slope be shallower for that epoch -- as is the case. These \alphaox\ slopes found are reasonable for an AGN of this UV luminosity (Vagnetti et al. 2013) and all fall within the span of values calculated by Grupe et al. 2010 (hereafter G10) for 92 Type I AGN. The \alphaox\ values indicate that \he0436 is behaving ``normally" (Gallo 2006) and extreme absorption or extreme relativistic effects are not likely to be present.


\subsection{Detailed SED Modeling Using \swift}
\label{SED}
Ideally, one would like to analyze a broadband spectral energy distribution (SED) that can be used in model fitting directly (e.g. Elvis et al. 1994). \swift\ UVOT data from epoch Sw1 were used for the broadband analysis of \he0436 as it provides data in five optical bands (as opposed to \xmm OM, which observed with only two filters). The XRT Sw1 data provided the 2 -- 10\keV spectrum and the 70-mo BAT data were included for the hard X-ray band coverage. 

Following the procedures outlined in Vasudevan \& Fabian (2009) and then expanded upon by Vasudevan et al. (2009) (hereafter V09), a basic model composed of a broken power law and a black body was applied to the SED. The break radius of the broken power law, \emph{R}$_{\rm{br}}$, is defined as the radius at which the photon index transitions from its inner value, $\Gamma_{1}$, to outer, $\Gamma_{2}$, in units of \emph{R}$_{\rm{g}}$. Parameters defined in the black body accretion disk model included the disk inner radius (\emph{R}$_{\rm{in}}$), the disk normalization (\emph{K}), and the maximum temperature of the accretion disk (\emph{T}$_{\rm{max}}$). \emph{T}$_{\rm{max}}$ was allowed to vary and the inner radius was set to 6\Rg, the innermost stable circular orbit (ISCO) of a Schwarzschild black hole. The normalization is defined as \emph{K} = $\frac{M^{2}_{BH}\thinspace cos\thinspace\emph{i}}{D^{2}_{\rm{L}}\thinspace \beta^4}$ where \emph{i} is the inclination of the disk, \emph{D}$_{\rm{L}}$ the luminosity distance to the source in kpc, and $\beta$ the spectral hardening ratio. The black hole mass for \he0436 was adopted from G10 as \Mbh\ = 5.9$\times10^7$\Msun\ and a luminosity distance of 233.1\Mpc\ was used.\footnote{http://www.astro.ucla.edu/~wright/CosmoCalc.html}  V09 demonstrated that cos\thinspace\emph{i} and $\beta$ could be set to 1 for type-I AGN, and thus the value for \emph{K} was calculated and fixed in the initial fits. 

The assumption regarding $\beta$ is reasonable due to the nature of the analysis and the \textsc{diskpn} model: $\beta$ is the $T_{color}$ / $T_{effective}$ ratio that is designed to correct flux in the far-UV band of a spectrum in order to preserve the total integrated bolometric luminosity (Shimura \& Takahara 1995). In the model \textsc{diskpn}, a decrease in $\beta$ will boost flux across all wavebands, not just in the far-UV. In addition, as only near-UV data were used in the SED fitting, no color correction is ultimately required. As for inclination or cos\thinspace\emph{i}, its influence on the overall normalization is minimal compared to both \Mbh\ and \emph{D}$_{\rm{L}}$, which are both raised to the second power. In general, a change in the normalization parameter, \emph{K}, \emph{overall} does not alter the spectral shape of the component significantly, but rather adjusts the peak temperature of the (estimated) black body in order to better fit the data.

Assuming a Comptonization origin for the primary X-rays, the low-energy break in the power law corresponded to \emph{T}$_{\rm{max}}$ of the disk component. Galactic absorption was set to $N_{\rm{H}}$ = 1.03$\times$$10^{20}$\thinspace$\rm{cm^{-2}}$ and intrinsic extinction was considered in various ways, as will be described below. Optical filters (V, B, and U) were not used for fitting as host galaxy contamination was not accounted for; the optical data were plotted simply for comparison. The XRT data were also fit over a reduced range of 2 -- 10\keV. This limit was placed to ensure the primary X-ray emission was fit rather than the soft excess which may originate from a different component. The BAT data were not fit at all and were simply overplotted for visual comparison.

The first model tested (Model A) described the simplest case where only Galactic extinction modified the intrinsic broadband spectrum. This scenario did not describe the data well (\chisq\ = 102 / 30 d.o.f), overestimating the UV spectral slope significantly. Galactic extinction alone can not explain these data. Host galaxy extinction was then added to the model and allowed to vary (Model B). The fit improved by \delchi = 71 for an additional free parameter. The extinction that described the SED best was \emph{E$_{\rm{B-V}}$} = 0.095$\pm$0.011. This scenario describes the broadband spectra very well, however it seems contradictory with the X-ray analysis performed previously. Assuming the estimated extinction, the corresponding column density can be calculated using the relationship outlined by Predehl \& Schmidt (1995): \nh\ = $5.3\times10^{21}$\thinspace{\rm{cm$^{-2}$}}\thinspace{\emph{E$_{\rm{B-V}}$}}. For\ \emph{E$_{\rm{B-V}}$} ranging from 0.084 -- 0.105, \nh\ = (5.04$\pm$0.58)$\times10^{20}$\thinspace{\rm{cm$^{-2}$}}, which is at least $50\times$ the upper limit measured from the X-ray spectra (see Section\thinspace\ref{firstmodeling}). Moreover, G10 measures extinction from the Balmer decrement and finds \emph{E$_{\rm{B-V}}$} = 0.04, which is much lower than the intrinsic extinction estimated in Model B. 

Next, the intrinsic extinction measured from the Balmer decrement was adopted and the \emph{K} was allowed to vary (Model C). The best value was \emph{K} = (52.16$\pm$14.30) $\times10^{-4}$\thinspace{\Msun$^2\kpc^{-2}$} and \emph{L$_{0.001-100\keV}$} = 6.47$\times10^{44}$\ergps, giving an Eddington luminosity ratio of \emph{L$_{\rm{Bol}}$} / \Ledd = 0.09. Model C provides a better fit than Models A and B, and the treatment of the extinction is physically motivated by the measured Balmer decrement.

The final possibility tested is that the SED can be entirely described by allowing only the disk normalization to vary with only Galactic extinction (Model D). The fit statistic was the comparable to Model C and \emph{K} = (107.54$\pm$25.77)$\times10^4$\thinspace{\Msun$^2\kpc^{-2}$}. In this case, \emph{L$_{0.001-100\keV}$} = 4.99$\times10^{44}$\ergps and so \emph{L$_{\rm{Bol}}$} / \Ledd = 0.07. 

A summary of the model variations with their respective parameters and Eddington ratios are listed in Table \ref{SEDmodels}. The SED models with high extinction values seem at odds with the initial X-ray analysis, extinction measured from the Balmer decrement (G10), and the UV spectral analysis of Halpern \& Marshall (1996). Alternatively, the SED can also be explained if we assume the source is not substantially reddened, but that the ratio (\Mbh\ / $D_{\rm{L}}$)$^{2}$ is uncertain by a factor of 6 -- 10.  This is perhaps not unreasonable.  The black hole mass for \he0436\ was not from reverberation mapping, but rather based on a single-epoch mass estimate which could be uncertain by a factor of a few (Vestergaard \& Peterson 2006).

\begin{figure}
   \centering         
   {\scalebox{0.32}{\includegraphics[angle=270, trim= 1cm 0cm 0.5cm 0.5cm, clip=true]{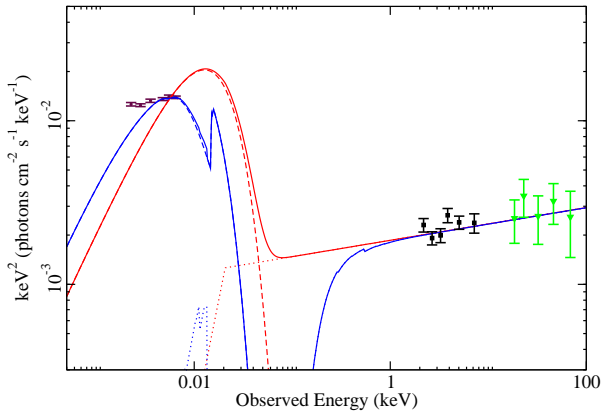}}} 
   \caption{An SED comparison of Model C with intrinsic extinction determined by the Balmer decrement and the disk parameter \emph{K} allowed to vary. The solid lines represent the total model in the source (red) and observed (blue) reference frames. The dashed line represents the disk component while dotted shows the primary X-ray component. Only the XRT data (black squares) from 2.0 -- 10.0\keV\ and the UV data of the UVOT (tiny maroon circles) were fit, the $70$-month \swift BAT data (green triangles) were overplotted for visual inspection. Only the XRT data that were fit (i.e. between 2 -- 10\keV) were included for clarity.}
   \label{SEDwBAT}
\end{figure}

The \swift BAT data were included for visual inspection of the high energy trend, however the BAT data were not fitted. Although the uncertainty in the BAT data allow for a wide spread, the extrapolated models were all consistent with these data. Model C (i.e. Balmer decrement based intrinsic extinction and \emph{K} allowed to vary) is shown for both the observer and source frame (see Fig.\thinspace\ref{SEDwBAT}).

The Eddington ratios determined from the \swift SED analysis agree with the value of 0.08 found by G10 for this object. An Eddington ratio around 0.08 is not unusual, although it does fall on the lower end of the probability distribution described by V09 in their analysis of 26 low-\emph{z} AGN (see V09 Fig. 15, lefthand panel). A lower Eddington ratio implies \he0436 is not a rapid accreter. 


\section{Detailed X-ray Spectral Analysis}
\label{XraySpec}
The first glance at the broadband X-ray spectra of \he0436 showed the presence of multiple components and the nature of the excess residuals above $\sim 6\keV$ remains unclear. Three scenarios are considered to describe the X-ray spectra of \he0436 in a self-consistent manner: partial covering absorption, blurred reflection, and soft Comptonization.


\subsection{Partial Covering Absorption}
\label{absorption}
Turner \& Miller (2009) explain in detail how gas with a range of column densities can influence the primary component of an AGN. As absorbers move in the field of view, some of the intrinsic radiation is both scattered and absorbed. Both changes in the covering fraction and column density can create spectral variability over a range of timescales. A strong soft excess can appear at lower energies and even the \feka\ line can be broadened by absorption with the red shoulder being enhanced by attenuation rather than by relativistic effects.  A thorough attempt was made to fit the X-ray spectrum of \he0436 with absorption in an effort to determine the origin of the X-ray emission in this object.

The model \textsc{zpcfabs} was used to begin testing for possible neutral absorption signatures. A single, neutral absorber was applied to a simple power law and fit to both epochs of the merged MOS data. Scenarios were tested in which only the covering fraction remained free to vary between the epochs (i.e. variability attributed to absorbers) and also having only the primary component vary instead (i.e. variability attributed to primary emitter). The overall goodness-of-fit remained poor in both cases. Double neutral absorption was an improvement, though the model continued to underestimate the data at higher energies. This scenario called for the two absorbers to have very different densities (see Table \ref{AbsTable}) and could potentially describe a single absorber with a density gradient along the line of sight, rather than two completely independent bodies. The column densities remain consistent in the two epochs and so would not alone explain any spectral variability seen over longer timescales. Allowing only the primary emitter to vary instead did not improve the fit quality significantly. Apparently, if there is any variability in the primary emitter on long timescales, it is subtle.


\begin{table*}
\caption{The best spectral absorption models in the 0.5 -- 10.0\keV\ band. Models are listed in Column 1 and their components in Column 2. Model parameters are in Column 3 and their values in Columns 4 and 5 for XMM1 and XMM2, respectively. Parameters linked across the epochs are denoted by dots.}
\centering
\scalebox{1.0}{
	\begin{tabular}{ccccc}
	\\
	\hline
	(1) & (2) & (3) & (4) & (5) \\
	\bf{Model} & \bf{Model Component} & \bf{Model Parameter} & \bf{XMM1} & \bf{XMM2} \\\hline
	\\
	 double neutral & absorber 1 & $N_{H}$ (\pscm) & $26.30^{+3.27}_{-2.94}$x$10^{22}$ & \ldots \\
	 absorption & & \emph{CF} & 0.61$\pm$0.02 & 0.56$\pm$0.02 \\
	 & absorber 2 & $N_{H}$ (\pscm) & $1.33^{+0.10}_{-0.09}$x$10^{22}$ & \ldots \\
	 & & \emph{CF} & 0.48$\pm$0.02 & 0.42$\pm$0.02 \\
	  & primary flux & $\Gamma$ & 2.66$\pm$0.03 & \ldots \\
	 & fit quality & \redchi/d.o.f. & 1.11/2030 & \\
	 \\
	 double ionized & absorber 1 & $N_{H}$ (\pscm) & $44.62^{+5.46}_{-6.27}$x$10^{22}$ & \ldots \\
	 absorption & & log($\xi$)\thinspace\small{[\ergcmps]} & $1.59^{+0.34}_{-0.30}$ & \ldots \\
	 & & \emph{CF} & $0.67^{+0.02}_{-0.03}$ & 0.63$\pm$0.03 \\
	 & absorber 2 & $N_{H}$ (\pscm) & $1.75^{+0.14}_{-0.13}$x$10^{22}$ & \ldots \\
	 & & log($\xi$)\thinspace\small{[\ergcmps]} & $0.48^{+0.13}_{-0.10}$ & \ldots \\
	 & & \emph{CF} & 0.52$\pm$0.03 & 0.44$\pm$0.04 \\
	  & primary flux & $\Gamma$ & 2.63$\pm$0.05 & \ldots \\
	 & fit quality & \redchi/d.o.f. & 1.06 / 2028 & \\
	\hline
	\end{tabular}
	}
\label{AbsTable}
\end{table*}

\subsection{Ionized Partial Covering}
\label{iabs}
Next, a fit with a single ionized absorber was attempted. The ionized absorber was modeled by \textsc{zxipcf}. Once again, all parameters except for covering fraction were linked at the epochs in an effort to describe the long-term variability with changes in an absorber. There were residuals to the model and the fit quality was poor (\chisq\ / d.o.f. = 2415 / 2032). The covering fraction was not significantly different from one epoch to the next.

If the photon index of the primary component is allowed to vary while the absorber is frozen and linked across the epochs, the fit improves by \delchi\ =  213 for two additional free parameters, however some positive residual features remain. The photon index in XMM2 increases, going from $\Gamma$ = 2.16$\pm$0.01 to $\Gamma$ = 2.24$\pm$0.01, consistent with the trend found in the toy model fits. It appears that the model with a single ionized absorber detects a slight, long-term variability from the primary emitter while the model with a single neutral absorber does not.

\begin{figure}
   \centering         
   {\scalebox{0.33}{
   	\includegraphics[angle=270, trim= 1cm 0cm 0.5cm 0.5cm, clip=true]{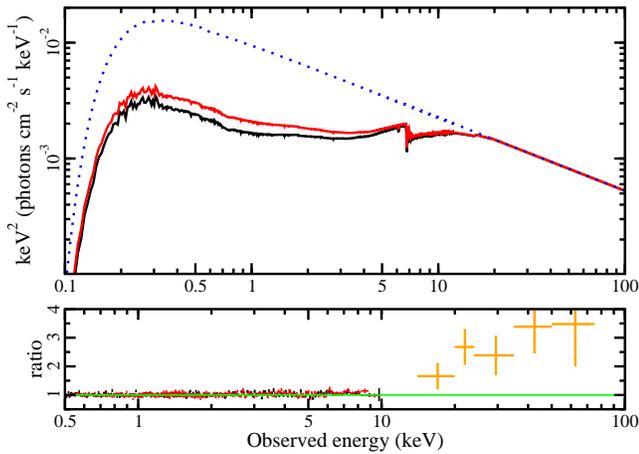}}} 
   \caption{The ionized partial covering model (Table \ref{AbsTable}) and its relation to the MOS spectra. A pair of ionized absorbers were fit from 0.5 -- 10.0\keV\ with the covering fraction as the only parameter varying at both epochs. The XMM1 (black) and XMM2 (red) data are shown with the constant intrinsic power law determined by the model (dotted blue) in the top panel. The model was extrapolated above 10\keV\* for visual comparison with $70$-month \swift\ BAT data (yellow) and the corresponding residuals are shown (bottom panel).}
   \label{iabs_resid}
\end{figure}

A second ionized absorber was added to the model and the parameters reset so that only the covering fractions were unlinked across the epochs. The fit was significantly better than that of the single absorber (\chisq / d.o.f. = 2150 / 2028) and no residual features remained (see Fig. \ref{iabs_resid}), although the covering fractions for both absorbing bodies remained similar at both epochs (see Table \ref{AbsTable}). Thus, as a final exercise, one absorber was kept constant throughout the epochs while the other was allowed to vary in \nh\ and $\xi$, with $\Gamma$ free to vary as well. The goodness-of-fit remained unchanged and the parameters that were allowed to be independent in XMM2 did not significantly change from their XMM1 values.

None of the ionized absorption model variations defined the source of long-term variability uniquely, however, the scenario in which only the covering fraction varied between epochs did result in the best goodness-of-fit. The most satisfying absorption model overall describes two ionized absorbers, the first more ionized and with a higher column density than the second, with both having comparable covering fractions. Both absorbers are found to have redshifts comparable to the source that remained steady across the epochs. 

The ionized partial covering model was extrapolated to higher energies and visually compared to the \swift BAT data. Model residuals show that, despite the partial covering scenario's ability to describe the 0.5 -- 10\keV spectra well, it under-predicts the X-ray flux above 10\keV\* (Fig. \ref{iabs_resid}, residuals). However, photoelectric absorption will also produce its own reprocessed emission in addition to the intrinsic radiation it attenuates (e.g. Turner \& Miller 2009). Reprocessed emission from absorption would affect both the hard X-ray flux and flux of the \feka\* line.

Following Reynolds \et (2004) we estimate the strength of an \feka\ emission line that would be produced from the denser of the two absorbers in the partial covering model.  Assuming the number of ionising photons between 7 -- 20\keV\ that are reemitted in the neutral iron line at $6.4\keV$ is governed by the fluorescent yield of neutral iron (0.347) we estimate the flux in the iron line should be approximately 2.70$\times10^{-13}$\ergpcmsps\ for XMM1, 2.55$\times10^{-13}$ \ergpcmsps\ for XMM2. 
Such a narrow line would have an equivalent width of  $EW\approx$  210\eV\ and 200\eV, in the spectrum of XMM1 and XMM2, respectively. The strength of the narrow \feka\ emission line in the spectra of \he0436\ is only about $40$~\eV, much weaker than expected if arising from reprocessing, but we note that various effect such as Compton-thick lines-of-sight could generated weaker lines than prediced (e.g. Yaqoob \et 2010).

Finally, an unblurred reflection component {\sc reflionx} (Ross \& Fabian 2005) was added to the absorption model to model for any narrow \feka\ emission, but also to adjust the residuals above $10\keV$ seen in Fig.~\ref{iabs_resid}. The modification above $10\keV$ was negligible, which is expected given the steep photon index required to fit the spectra below $10\keV$.


\subsection{Blurred Reflection}
\label{sec:meanref}
The multi-epoch observations of \he0436 were next fit with a blurred reflection model. The primary X-ray source was modelled as a cutoff power law with the high-energy cutoff fixed at $300\keV$. The backscattered emission from the accretion disk, modelled with {\sc reflionx}, was modified by relativistic effects close to the black hole (Fabian et al. 1989). We adopted {\sc kdblur2} for the blurring kernel, which incorporates a twice broken power law for the emissivity profile. 

The disk inclination ($i$), iron abundance ($A_{Fe}$), and inner disk radius ($R_{in}$) are not expected to change over the observable time scales and were linked between epochs. The iron abundance was initially fixed to solar, the outer disk emissivity profile ($q_{out}$) was fixed to $3$, and the break radius ($R_b$) where the emissivity index changes from $q_{in}$ to $q_{out}$ was fixed to $R_b$ = 6\rg\* = 6 $\frac{GM}{c^2}$. At each epoch, the ionization parameter ($\xi=4\pi F/n$ where $n$ is the hydrogen number density and $F$ is the incident flux), normalization of the reflector, the incident power law photon index, and normalization were all free parameters.    

\begin{table*}
\caption{The best-fit blurred reflection model fit simultaneously to the \xmm merged MOS spectra. The model components and model parameters are listed in Columns 1, and 2, respectively. Columns 3 and 4 list the parameter values during the XMM1 epoch and XMM2 epoch, respectively. Parameters that are fixed are listed without errors and the dots denote those that are linked across the epochs. The reflection fraction ($\mathcal{R}$) is approximated as the ratio of the reflected flux over the power law flux in the 0.1-100\keV\* band. Fluxes are corrected for Galactic absorption and are reported in units of \ergpscmps.}
\centering
\scalebox{1.0}{
	\begin{tabular}{ccccc}                
	\hline
	(1) & (2) & (3) & (4)  \\
 	Model Component &  Model Parameter  &  XMM1 & XMM2 \\
       \hline
 	Incident & $\Gamma$ & 2.03$\pm$0.02 & $2.14^{+0.03}_{-0.07}$  \\
 	Continuum & $E_{\rm{cut}}$ (keV) & 300 & \ldots \\
	&$F_{0.1-100\thinspace\rm{keV}}$ & 1.35$\pm0.06\times10^{-11}$ & 1.66$\pm0.08\times10^{-11}$    \\
	Blurring & $q_{\rm{in}}$ & $6.0^{+2.0}_{-0.7}$ & $5.6\pm0.7$ \\
         & $R_{\rm{in}}$ (\rg) & $<1.80$ & \ldots \\
         & $R_{\rm{out}}$ (\rg)   & 400 & \ldots \\
         & $R_{\rm{b}}$ (\rg) & 6 & \ldots \\
         & $q_{\rm{out}}$ & 3 & \ldots \\
         & $i$ (\deg)  & $45^{+8}_{-7}$  & \ldots \\
  	Reflection & $\xi$ (\erg\cmps) & $150^{+320}_{-48}$  & $56^{+41}_{-11}$ \\
         & $A_{\rm{Fe}}$ (solar) & $0.36^{+0.04}_{-0.16}$ & \ldots \\
         &$F_{0.1-100\thinspace\rm{keV}}$ & 4.90$\pm0.63\times10^{-12}$ & 4.79$\pm0.52\times10^{-12}$ \\
         & $\mathcal{R}$ & 0.36$\pm$0.05 & $0.29\pm0.03$   \\
 	Distant Reflector & $\xi$ (\erg\cmps)   & 1.0  &  \ldots \\
         & $A_{\rm{Fe}}$ (solar) & 1.0 & \ldots \\
         & $\Gamma$ & 1.9 & \ldots  \\  
         &$F_{0.1-100\thinspace\rm{keV}}$ & $1.62^{+0.52}_{-0.20}$  $\times10^{-12}$ & \ldots \\
         Fit Quality & \chidof & 1.06 / 2024    &  \\
         \hline
\label{tab:Meanfit}
\end{tabular}
}
\end{table*}

\begin{figure*}
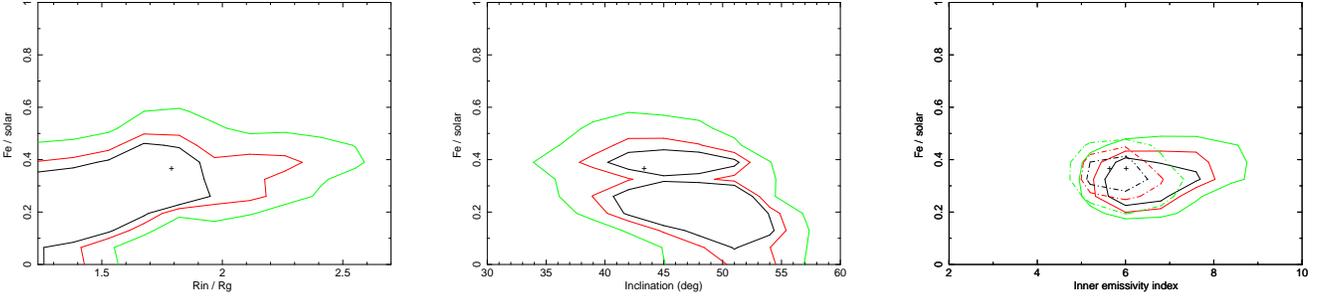

\centering
\scalebox{0.22}{\includegraphics[angle=270]{C_fe_rin.ps}}\hfill
\scalebox{0.22}{\includegraphics[angle=270]{C_fe_inc.ps}}\hfill
\scalebox{0.22}{\includegraphics[angle=270]{C_fe_QD.ps}}
\caption{Contours are calculated to examine the dependency of the measured iron abundance with various other parameters. The iron abundance is compared to the inner disk radius (left), disk inclination (center), and inner emissivity profile (right). In the right panel the solid and dashed contours are for data set 1 and 2, respectively. 
The contours correspond to a 2.3, 4.61 and 9.21 delta fit statistic.}
\label{fig:contour}
\end{figure*}

The measurement of iron abundance is rather sensitive and degeneracies have been found with other fit parameters like emissivity, inclination, and black hole spin (or comparably, the inner disk radius). Contour plots were generated between various fit parameters to determine if the low iron abundance could be an artifact of the modelling, but in all cases the iron abundance remained consistently low (Fig.~\ref{fig:contour}).

A second {\sc reflionx} component that was not blurred and whose flux remains linked between epochs was also included to mimic emission from a distant reflector like the torus. The ionization parameter of this distant reflector was fixed at $\xi=1\ergcmps$ and the iron abundance set to the solar value. The photon index of the power law source illuminating the torus is fixed at $\Gamma=1.9$, the canonical value for AGN (e.g. Nandra \& Pounds 1994). There is no reason to assume that the radiation incident on the distant reflector will be the same as that from the primary component and if a value of $\Gamma=2.0$ is used instead it does not significantly alter the fit statistic.

The described model produced a reasonable fit to both spectra (\chidof=1.08/2025), but left excess residuals above $5\keV$. Permitting $q_{out}$ and $R_b$ to vary at each epoch did not substantially improve the fits, but a significant improvement (\delchi=33 for 1 additional fit parameter) was found if the iron abundance was allowed to vary. The resulting fit was good (\chidof=1.06/2024; Table~\ref{tab:Meanfit}), but noteworthy was the low, sub-solar iron abundance measurement ($A_{Fe}$= $0.36^{+0.04}_{-0.16}$).

Further examination to determine if the low iron abundance was a result of the fit procedure were carried out. The iron abundance of the distant reflector was linked with the inner, blurred reflector to determine if enhancing the iron ``signal'' could result in a more typical measurement. This fit was an improvement over the initial fit, with the abundance fixed at solar (\delchi=25 for 1 additional fit parameter), but the resulting measurement was comparable to the measurement from the blurred component alone ($A_{Fe}$ = $0.33^{+0.14}_{-0.17}$).

Strong degeneracy has been noted between the black hole spin parameter ($a$) and the iron abundance (e.g. Reynolds 2013). The weak reflection features and moderate signal-to-noise of the data do not justify a thorough examination into black hole spin. However, the possibility that the low iron abundance may be rectified with treatment of the black hole spin should be considered. With this in mind, the blurring kernel {\sc kdblur2} was replaced with {\sc relconv} (Dauser et al. 2010) which includes (retrograde and prograde) black hole spin as a measured parameter. The fit quality was comparable to that achieved with {\sc kdblur2} (\chidof=1.06/2024). The spin parameter was high ($a\approx 0.96$) and was consistent with the value of the inner disk radius reported in Table~\ref{tab:Meanfit}. The iron abundance continued to remain low ($A_{Fe} \approx0.28$). Forcing the model to accommodate a retrograde spin ($ -0.998 \le a \le 0$) only resulted in a poorer fit (\chidof=1.11/2024) and did little to enhance the iron abundance ($A_{Fe} \approx0.56$).

Further consideration was to adopt a different reflection code. The reflection model {\sc reflionx} was replaced with {\sc xillver} (Garcia \et 2013).   There are several differences between {\sc xillver} and {\sc reflionx} including {\sc xillver}'s ability to calculate reflected intensity for each photon energy, position, and viewing angle whereas {\sc reflionx} uses an angle-average integration for reflected intensity. Primarily, we consider if these differences could be attributed to the low iron abundnace in \he0436.  Substituting {\sc reflionx} for {\sc xillver} in the original model (Table~\ref{tab:Meanfit}) resulted in an equally good fit (\chidof=1.07/2024) with comparable parameters measurements. We note the iron abundance boundary value in {\sc xillver} is $0.5$, which makes it impossible to explore values as low as we found with {\sc reflionx}. Nonetheless, the best-fit iron abundance pegs at the lower limit of the model ($A_{Fe} $=$ 0.50^{+0.23}_{-0}$).

\begin{figure}
\rotatebox{270}
{\scalebox{0.32}{\includegraphics{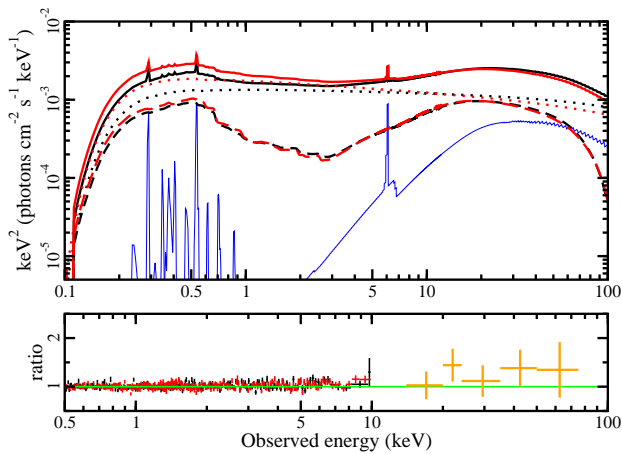}}}
\caption{The blurred reflection model for the two epochs (top panel). The primary X-ray component is dotted while the blurred reflector is dashed. The distant reflector (same in both epochs) is solid blue. The total summed model is solid black and red, for XMM1 and XMM2 respectively. The model was extrapolated above 10\keV\* for visual inspection with the BAT data (yellow) and the residuals (model/data) are shown (bottom panel).}
\label{fig:ratio}
\end{figure}


\subsection{Soft Comptonization}
\label{SoftCompt}
Both the absorption and reflection scenarios addressed in the previous sections fit the soft-excess of \he0436 well and it is statistically impossible to distinguish between the two scenarios without further analysis. A third possibility is that the soft-excess could be emerging from a second Comptonizing region.

Observations of binary black holes have shown that spectral fitting can require two Comptonizing regions (Kubota \& Done 2004 and Done et al. 2012, hereafter D12). While there are several possibilities for the source of the second softer Compton up-scattered photons, D12 assume such soft Comptonization is originating from the accretion disk itself. The authors postulate that disk structures could deviate from the standard Shakura-Sunyaev model at inner radii, as has been proposed for some of the more extreme Seyfert galaxies like Narrow-Line Seyfert 1s (NLS1s). In particular they consider enhanced dissipation of disk outer layers that would create a Compton up-scattered flux in the soft X-ray band.

As an initial investigation into a double-Comptonization scenario, the merged MOS spectra were fit from 0.5 -- 10.0\keV\ with a basic power law + Comptonization model with the second Comptonization component provided by {\sc compTT} (Titarchuk 1994). The seed photons were assumed to be supplied by the intrinsic UV disk radiation below the observed bandpass and both the seed photon temperature and the temperature of the second Comptonizing region were allowed to vary. This lone double-Comptonization model did not fit well (\chisq\ / d.o.f. = 2319 / 2030) and there were clear residuals around 6.4\keV. When a distant reflector was added, the fit improved by \delchi\ = 99 for two additional free parameters, but the new model did not account for all of the iron line flux. If a broad line was added in the form of blurred reflection, the spectra were well-modeled in the 5 -- 8\keV\ range, however the normalization of the second Compton component dropped by an order of magnitude and the optical depth reduced from around 2\keV\ to ${\tau\sim}$\thinspace0.09 as blurred reflection now accounted for the soft band. A simple double-Comptonization scenario appears to model the soft-excess appropriately, but does not fit some of the curvature between 0.5 -- 10.0\keV.

\begin{figure}
\rotatebox{270}
{\scalebox{0.32}{\includegraphics{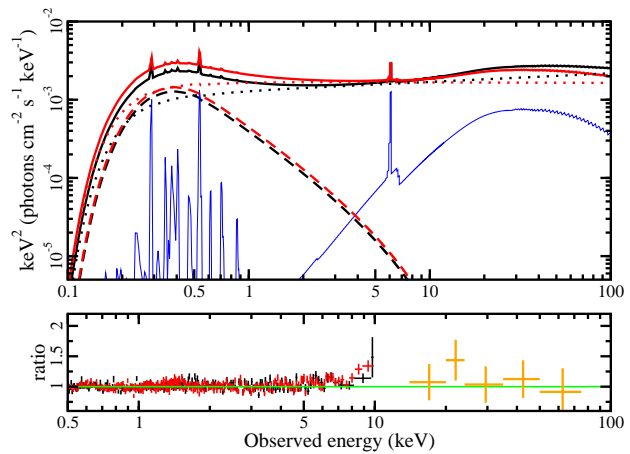}}}
\caption{The double-Comptonization model for the two epochs (top panel). The primary X-ray component is dotted while the second Compton component is dashed. The distant reflector (same in both epochs) is solid blue. The total summed model is solid black and red, for XMM1 and XMM2 respectively. The model was extrapolated above 10\keV\* for visual inspection with the BAT data (yellow) and the residuals (model/data) are shown (bottom panel).}
\label{dCmod}
\end{figure}

In an attempt to model the full 0.5 -- 10.0\keV\ spectra more physically, the data were fit with {\sc optxagnf} (D12). Black hole mass and luminosity distance were frozen at their known values of 5.9$\times$$10^{7}$\Msun\* and 233.1\Mpc, respectively. The Eddington fraction was set at log(\emph{L}/$L_{Edd}$) = -1.05 as determined by the SED fitting, a value that --- as mentioned previously --- agrees with the fraction found by G10. Black hole spin was set to \emph{a} = 0.5 to start and the normalization was fixed at unity as per model instructions\protect\footnote{https://heasarc.gsfc.nasa.gov/xanadu/xspec/models/optxagn.html}. For the initial fitting, only the power law parameters were allowed to vary independently across the epochs and all other parameters were frozen. 

Parameters were systematically thawed and unlinked to gradually increase complexity as needed to improve the fit. The parameters $f_{\rm{pl}}$,\thinspace\Rcor,\thinspace\emph{k}$T_{\rm{e}}$, and $\tau_{e}$ remained consistent between the epochs when allowed to vary and so were kept linked. Although it is reasonable to allow these components to vary between epochs, they did not improve the quality of the fit if unlinked and so, after an initial test of allowing them free to vary at both epochs, they were re-linked.

The initial fit was very poor (\chisq\ / d.o.f. = 2475 / 2032) with a clear iron line residual and other residual features at about \emph{E} $\leq$ 1\keV. The fit proved difficult to constrain when the spin parameter was allowed to vary freely, so the fit quality was instead tested at different fixed values of \emph{a}. A zero spin only worsened the fit (\delchi\ = 148), while testing a maximum spin value of \emph{a} = 0.998 improved the fit by \delchi\ = 303 for no additional free parameters. Fixing the spin at a moderately high value of \emph{a} = 0.8 improved the goodness-of-fit even further, with \delchi\ = 12 for the same number of free parameters, and \emph{a} was kept frozen at this value. If the Eddington ratio is allowed to vary with spin, neither parameter changes significantly from their best-fit values and the fit statistic remains comparable.

To improve the fit, a distant reflector was added in the same manner as Section \ref{sec:meanref}. The distant reflector improved the fit significantly, \delchi\ = 101 for one additional free parameter, and the iron line residuals were resolved. Curiously, the data remained overestimated around 0.8\keV\* until an absorption edge was included at the redshift of the source, with both its energy and optical depth free to vary. The absorption feature was found to have an energy of \emph{E}\thinspace$\sim$\thinspace0.77\keV\* and $\tau\thinspace\sim\thinspace$0.09 -- consistent with O\thinspace\textsc{vii}, though the accompanying O\thinspace\textsc{viii} absorption edge was not detected.  We noted this is the only model that shows marked improvement with the addition of an absorption edge. The addition of both a distant reflector and an absorption edge provided the best description of the data in the soft Comptonization scenario (\chisq\ / d.o.f. = 2211 / 2029) and the model that most satisfies the data is summarized in Table \ref{BestSC}.

\begin{table}
\centering
\advance\leftskip-0.8cm  
\caption{Summary for the best soft Comptonization model. All parameters are frozen with the exception of: \emph{R$_{cor}$}, \emph{kT$_{e}$, $\tau$}, $\Gamma$, \emph{f$_{pl}$} and all parameters are linked across the epochs with the exception of $\Gamma$.}
\scalebox{.9}{
		\begin{tabular}{cccc}
		\hline
		(1) & (2) & (3) & (4) \\
		 \bf{Component} & \bf{Parameters} & \bf{XMM1} & \bf{XMM2} \\\hline
		 Distant Reflection & $A_{\rm{Fe}}$ (solar) & 1.0 & \ldots \\
		 & $\Gamma_{\rm{dr}}$ & 1.0 & \ldots \\
		 & $\xi$\thinspace\small{(\ergcmps)} & 1.0 & \ldots \\
		 & \emph{z} & 0.053 & \ldots \\
		 & $norm_{\rm{dr}}$ & $(1.56\pm0.35)\times10^{-5}$ & \ldots \\
		 \\
		 Absorption Edge & $E_{\rm{edge}}$ (keV) & $0.77^{+0.03}_{-0.02}$ & \ldots \\
		 & $\tau_{\rm{max,edge}}$ & $(9.13\pm0.02)\times10^{-2}$ & \ldots \\
		 \\
		 Soft Comptonization & \Mbh\ (\Msun) & 5.9$\times10^{7}$ &\ldots \\
		 & $D_{\rm{L}}$ (\Mpc) & 233.1 & \dots \\
		 & log(\emph{L} / \Ledd) & -1.05 & \ldots \\
		 & log\Rout & 3 & \dots \\
		 & \emph{a} & 0.8 & \ldots \\
		 & \Rcor\thinspace\small{(\Rg)} & $11.55^{+5.41}_{-1.50}$ & \ldots \\
		 & $kT_{\rm{e}}$\thinspace\small{(keV)} & $0.61^{+0.20}_{-0.11}$ & \ldots \\
		 & $\tau$ & $7.73^{+1.19}_{-1.27}$ & \ldots \\
		 & $\Gamma$ & $1.80^{+0.01}_{-0.02}$ & $1.98^{+0.03}_{-0.05}$ \\
		 & $f_{\rm{pl}}$ & 0.44$\pm$0.01 & \ldots \\
		 \redchi / d.o.f. & 1.09 / 2029\\
		 \hline
		\end{tabular}}
\label{BestSC}
\end{table}

\begin{figure}
   \centering          
    {\scalebox{0.3}{\includegraphics[angle=270, trim= 1cm 0cm 0.5cm 0cm, clip=true]{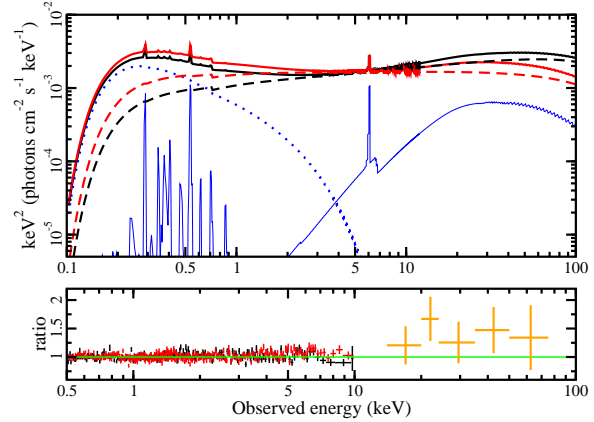}}} 
   \caption{The soft Comptonization model (Table \ref{BestSC}) is shown in the top panel. XMM1 is black and XMM2 is red. The hard power law component is shown as the dashed line. The soft Comptonization component that is common at both epochs is shown as a dotted blue curve. The distant reflect, also common at both epochs, is shown as the fine, solid, blue curve. The model was extrapolated $>$10\keV\* for comparison with the BAT data (yellow) and the corresponding residuals (data/model) are shown (bottom panel).}
   \label{SCBATPred}
\end{figure}

The model was extrapolated into the 10 -- 100\keV\* band and the BAT data was over plotted for comparison. The model components (top panel) and residuals that include the BAT data (bottom panel) are shown in Fig. \ref{SCBATPred}. The BAT data were only over plotted for comparison and were not included in the fit. The model slightly underestimates the fluxes seen above 10\keV.

Unlike the partial covering or blurred reflection models, the \textsc{optxagnf} model provides some information regarding the UV, disk dominated emission. Thus, it is advantageous to explore if the best soft Comptonization model is consistent with the \swift UVOT data. Simply extrapolating the model to the UV band shows a poor fit using the assumed black hole mass and luminosity distance, as did our initial SED fits (Section~\ref{SED}). Allow the black hole mass and distance to vary does not significantly improve the fit since the statistics are driven by the X-ray data, but it does suggest there could be a discrepancy of at least 2 in the \Mbh/$D_{\rm{L}}$ ratio comparable with our finding with the SED modelling. 

In an further attempt to fit the UV data, the parameters defining the disk component --- black hole mass, Eddington ratio, and spin --- were all allowed to vary. The fit is comparable to that of the X-ray only data (\chisq\ / d.o.f. = 2246 / 2029), however the model underestimated the UV flux. The black hole mass increased by $\sim$ 27\% while spin decreased by $\sim$ 20\%; the Eddington ration remained unchanged. If the remaining parameters -- those that describe the X-ray components and the absorption edge -- are free, the fit statistic does not change significantly and neither do the parameter values. Lastly, an absorber was added to the model to mimic intrinsic extinction in the UV.  This broadband soft Comptonization model with absorption fit well (\chisq\ / d.o.f. = 2213 / 2028), however it predicted an intrinsic extinction of $E_{B-V,int}$ = 0.055 which requires a column density of \nh\ = 2.99$\times10^{20}\pscm$: 29x higher than that predicted by the X-ray data (\nh $< 10^{19}\pscm$). Once again it appears that an absorption scenario is statistically acceptable for \he0436, albeit inconsistent with further scrutiny.


\begin{figure*}
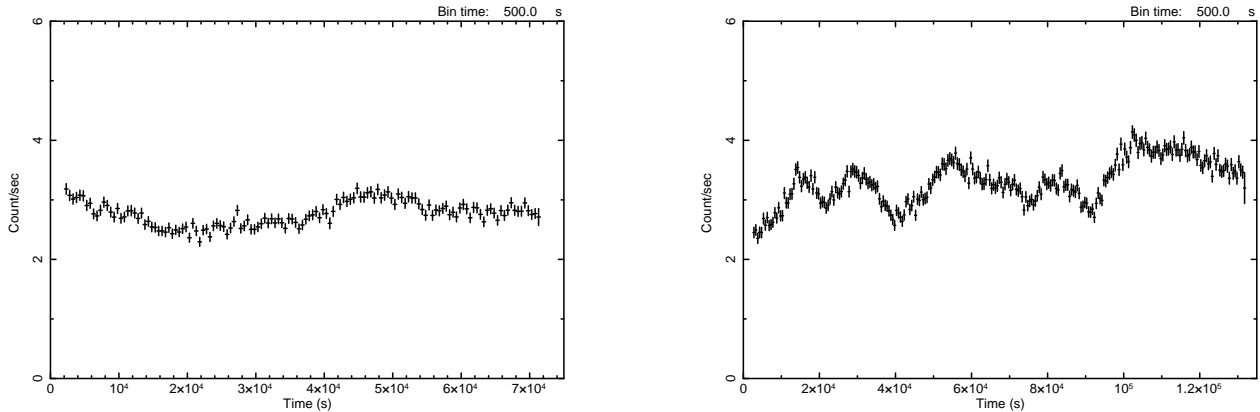

\begin{center}
\vspace{0.5in}
\begin{minipage}{0.48\linewidth}
	\scalebox{0.32}{\includegraphics[angle=270,trim= 0cm 0cm 0.7cm 0cm, clip=true]{XMM1_lc_bold.ps}}
\end{minipage}  \hfill
\begin{minipage}{0.48\linewidth}
	\scalebox{0.32}{\includegraphics[angle=270,trim= 0cm 0cm 0.7cm 0cm, clip=true]{XMM2_lc_bold.ps}}
\end{minipage}
\end{center}
\caption{MOS light curves for both XMM1 (left) and XMM2 (right) binned at 500\thinspace{s} each. XMM2 had nearly double the duration (note x-axes) and also displays more variability.  Zero marks the start of each exposure.}
\label{LCs}
\end{figure*}

\section{Timing Analysis}
\label{Timing}
Source photons were extracted to create light curves in a variety of energy bands between 0.2 -- 12.0\keV, including a broadband 0.2 -- 12.0\keV\ bin. The light curves from all EPIC instruments that collected data proved consistent in the XMM1 and XMM2. All subsequent timing analyses utilized MOS data only. MOS 1 and 2 light curves were combined at each epoch, taking care to match the observation start and stop times. The 0.2 -- 12.0\keV\ merged MOS light curve for each epoch is shown in Fig.\thinspace\ref{LCs}.

\subsection{Hardness Ratios}
\label{HRs}
Ratios of softer energy band count rates over those of a harder energy band were calculated for four energy band pairs between  0.2 -- 12.0\keV\ range. Bins were selected such that there were a comparable number of counts in each band. These hardness ratios (HR) were analyzed two ways at each epoch: flux-dependent HR; and as time-dependent, normalized HR.

The hardness as a function of count rate did not reveal any significant correlation at either epoch. This does not mean the spectra are not variable, but only that any variability is not significantly dependent on the specific flux state. The hardness as a function of time was similar: within the calculated uncertainties it appears there is little time-dependent variation in the soft bands over time. All HR trends for the X-ray data were nearly linear within uncertainties and so indicate little to no change in hardness ratio with respect to either flux or time. This does not mean that the spectral components are not variable, but rather that any variability that may be present occurs in such a way as to maintain a constant flux ratio between energy bins. A more thorough analysis is needed to further probe the temporal behavior in \he0436 and this is explained in the following sections.

\subsection{Fractional Variability}
\label{Fvar}
The fractional variability amplitude (Edelson et al. 2002), \Fvar, is a useful measurement used to quantify the intrinsic variability of a light curve while also accounting for uncertainty. This parameter compares the amplitude of any variations present in each energy band to determine if the amplitudes in one energy band are larger than those in another. The \Fvar\ thus compares the relative ``strengths" of observed variability across two energy bands, independent of time or flux. The \Fvar\ is defined as,
%
\begin{equation}
F_{\rm{var}} = \frac{1}{\langle X \rangle}\sqrt{S^2 - \langle \sigma^2_{\rm{err}} \rangle}
\label{FvarEq}
\end{equation}
%
where $\langle X \rangle$ is the mean count rate, $S^2$ the total variance of the light curve, and $\langle \sigma^2_{\rm{err}} \rangle$ the mean error squared. Uncertainties were determined following the procedure of Ponti et al. (2004). 

The \Fvar\ was calculated for both epochs using merged MOS light curves. The light curves were binned at 1000\thinspace{s} and a total of nine energy bins were used spanning 0.2 -- 10.0\keV; data above 10\keV\ were ignored due to the low signal-to-noise. XMM2 displays a higher \Fvar\ than XMM1 overall, which is apparent in the light curves (e.g. Fig.~\ref{LCs}).

\begin{figure}
   \centering    
   	{\scalebox{0.7}{\includegraphics[width=4.5in,trim= 0cm 0cm 0cm 4cm, clip=true]{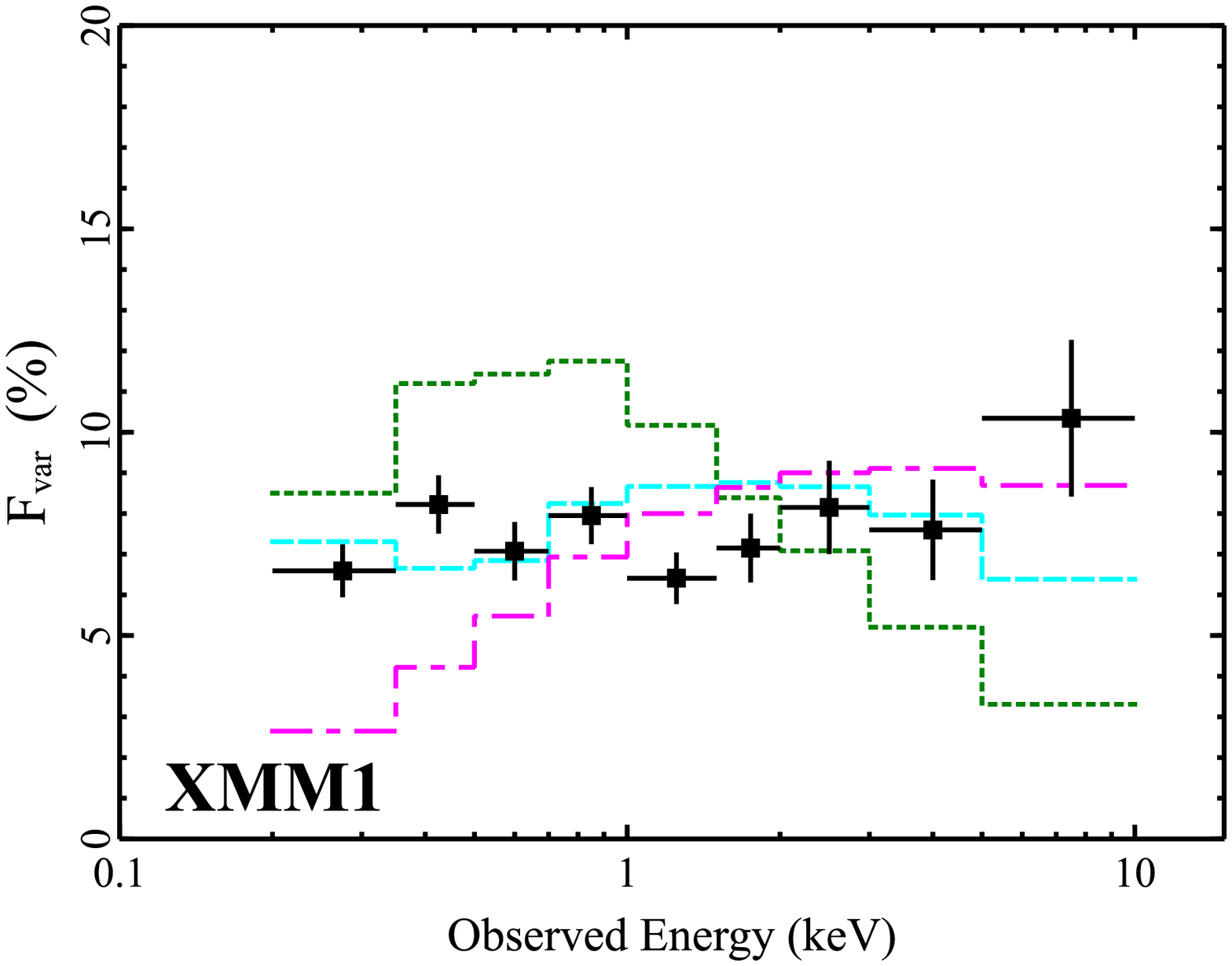}}}
	{\scalebox{0.7}{\includegraphics[width=4.5in,trim= 0cm 0cm 0cm 4cm, clip=true]{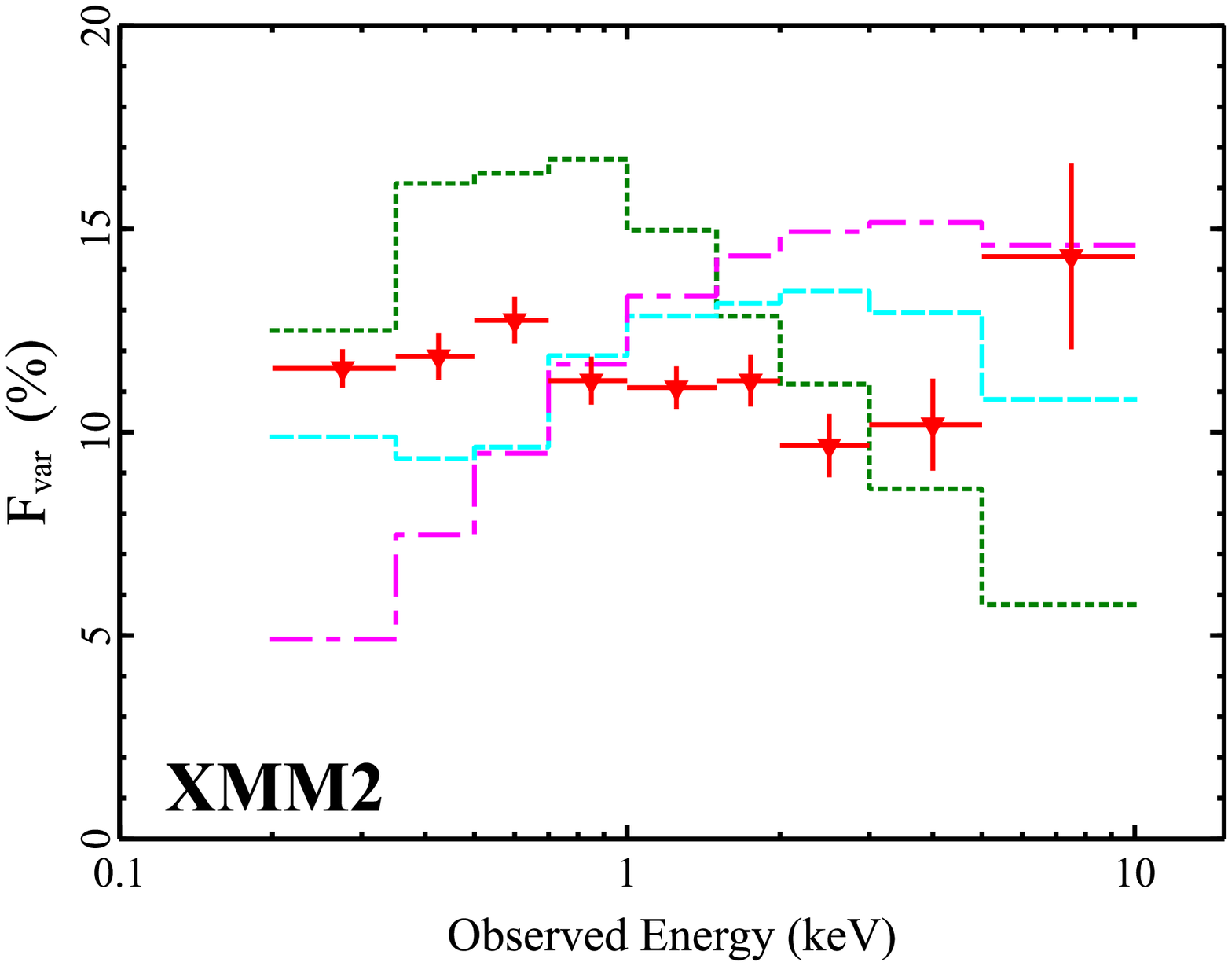}}}
   \caption{Fractional variability analysis with spectral model predictions. The fraction that the power law contributes to each energy band is overplotted on the \Fvar\ spectrum for both XMM1 (top panel) and XMM2 (bottom panel). The reflection (blue dashed line), absorption (green dotted line), and double-Comptonization (magenta dot-dashed line) model predictions are compared for each epoch.}
   \label{FvarMs}
\end{figure}

In terms of the blurred reflection model, variations in the primary component (i.e. power law) may solely account for the shape of the \Fvar\ spectrum. From the blurred reflection model found previously, the fraction that the power law contributes to each energy band was determined and overplotted on the \Fvar\ spectrum for both epochs (Fig. \ref{FvarMs}, dashed blue line). In the blurred reflection model it seems reasonable to attribute most of the rapid variability to changes in the power law normalization, specifically. In this very simplified scenario, the reflection model prediction disagrees with the data in a few bins (around 0.4\keV and between 1 -- 2\keV, for example). The best that can be said is that the overall model trend does agree with the linear shape of the fractional variability spectrum.

When the process is repeated for the partial covering absorption model, the predicted \Fvar\ profile is less constrained (Fig. \ref{FvarMs}, dotted green line). Testing the partial covering scenario is more challenging as the source of variability could be the absorber as well as the primary component. The same difficulty lies in modeling soft Comptonization: there could be many sources of variability. For completeness, a profile was created using the double-Comptonization model of Section \ref{SoftCompt}, with the hard Compton component assumed to be the sole source of variability (Fig. \ref{FvarMs}, dot-dashed magenta line).

Looking at the three model predictions for both epochs, it is clear that there must be more to the story. Perhaps only a fraction of the total power law flux is varying or perhaps an additional component contributes to the total variability. The \Fvar\ analysis indicates that multiple parameters need to vary if any model is to describe the spectral data.

\subsection{Flux-Resolved Spectra}
\label{FRS}
Flux-resolved spectra were made in an effort to resolve the mechanism(s) behind short-term variability in \he0436. As XMM2 was considerably longer and shows larger amplitude variations, data from the second epoch alone were used to create X-ray spectra with $<$\thinspace3\cps and $>\thinspace$3\cps.

A difference spectrum (high flux  -- low flux) was created to identify components that vary between the flux states. If changes between the flux levels were only due to changes in the normalization of the spectral components then a single power law should describe the difference spectrum well. Fitting the 3 -- 10\keV\ band with a power law ($\Gamma=2.00\pm0.16$) resulted in a good fit.  Extrapolating the power law to $0.5\keV$ reveal a strong excess toward lower energies.  Normalization changes may be responsible for the majority of variability in the broadband spectra, although changes in the shape of the spectral components at lower energies may be important. 

The two flux-resolved spectra were then fit with the models for partial covering absorption, blurred reflection, and Comptonization to examine the origin of the rapid variability. In the absorption model, changes in X-ray flux could be caused by either the primary emitter or by the absorbing material. Two sources of ionized absorption explain the X-ray spectra best (Section \ref{iabs}), one with higher column density than the other. Geometrically, these could represent two distinct absorbing materials or a single absorber with a density gradient. Ionization could also vary across a single larger body as the material closer to incident radiation would experience more illumination than that on the far side. As absorbing material moves, overlaps, or condenses/disperses a spectrum can change. The simplest scenario assumes variability caused only by a change in covering fraction, with all other properties of the absorbers remaining constant. 

\begin{figure}
\centering
{\scalebox{0.32}{\includegraphics[angle=270, trim= 1cm 0cm 0.5cm 0.5cm, clip=true]{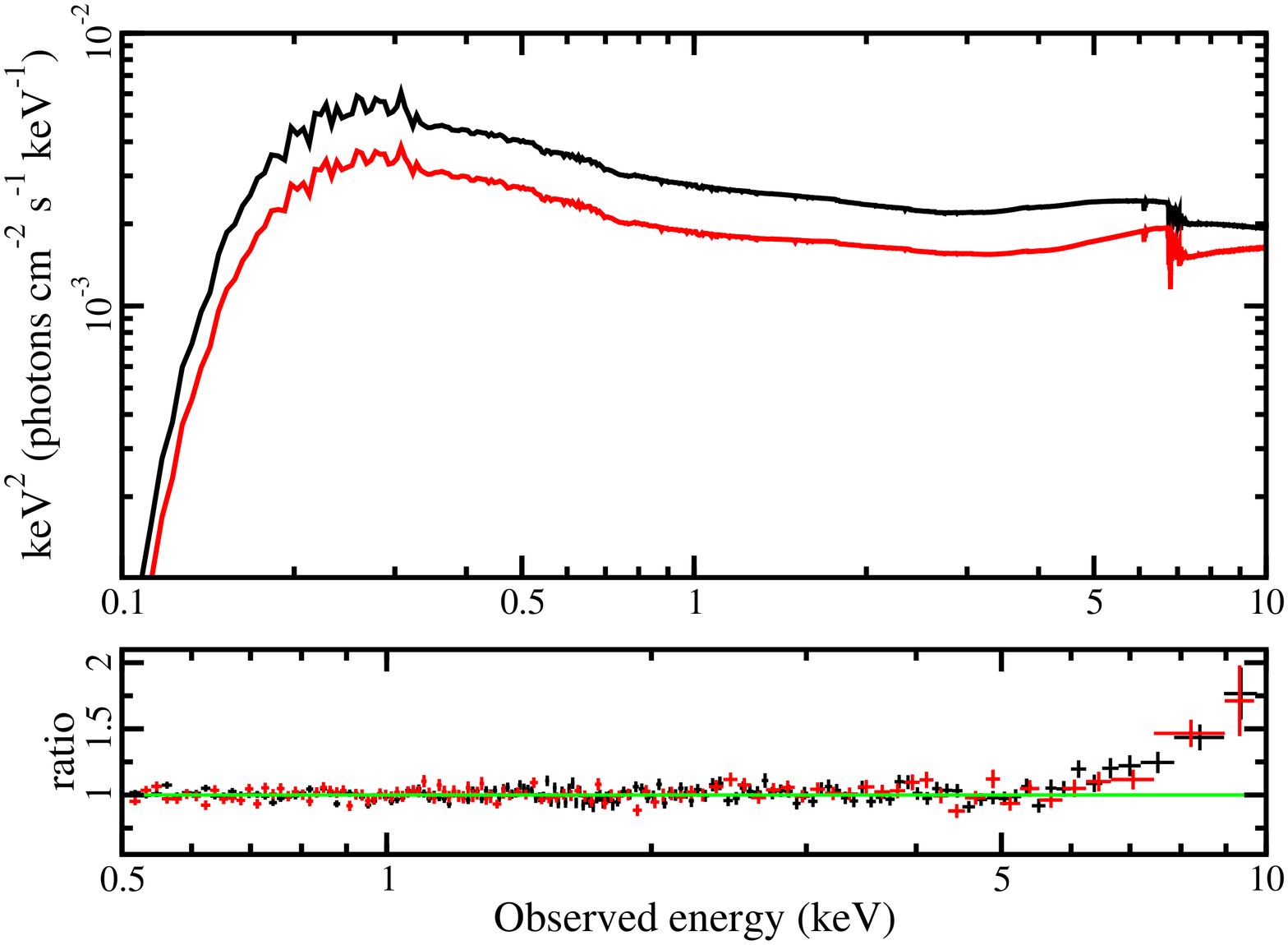}}}
{\scalebox{0.32}{\includegraphics[angle=270, trim= 1cm 0cm 0.5cm 0.5cm, clip=true]{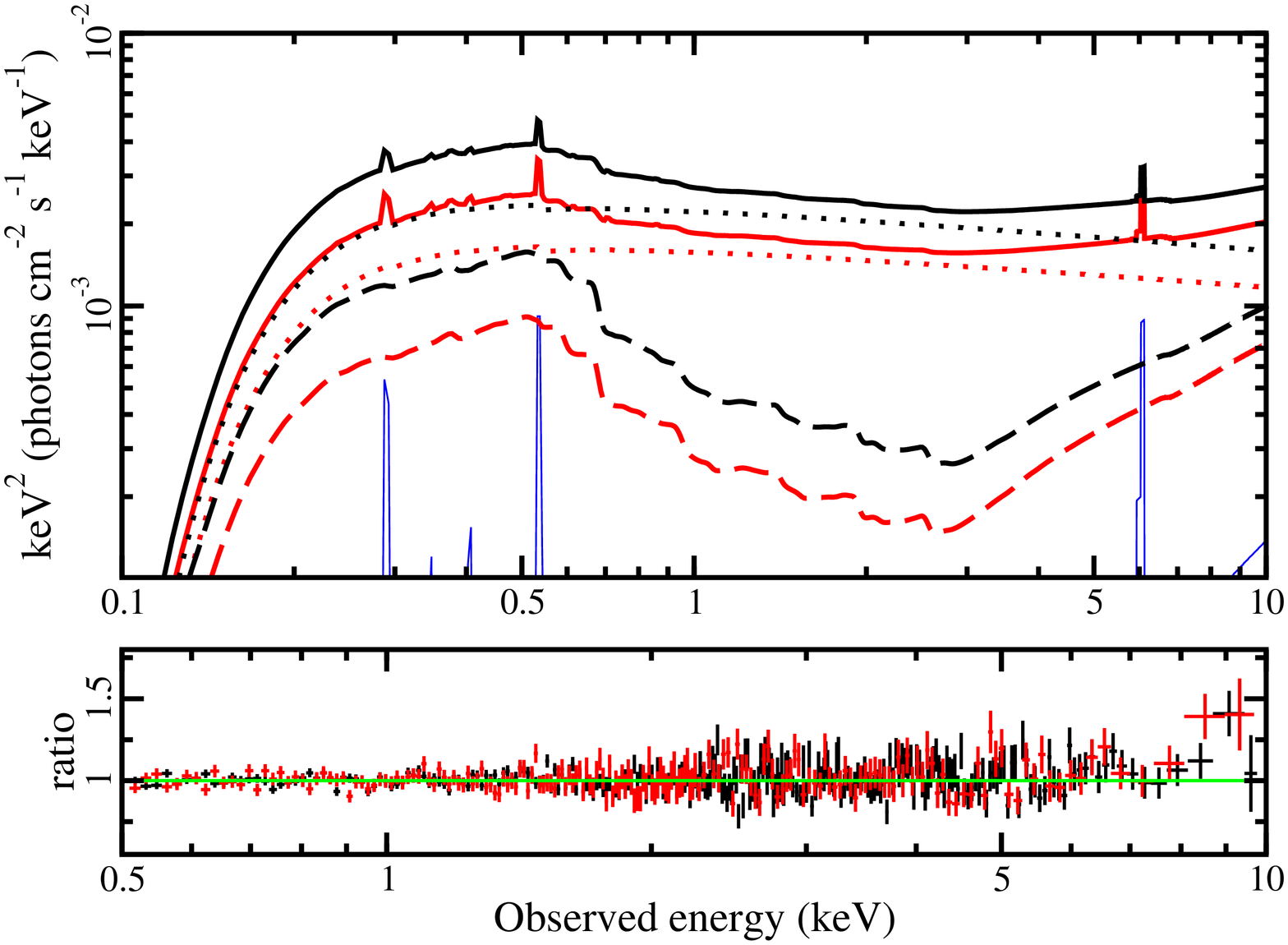}}}
{\scalebox{0.32}{\includegraphics[angle=270, trim= 1cm 0cm 0.5cm 0.5cm, clip=true]{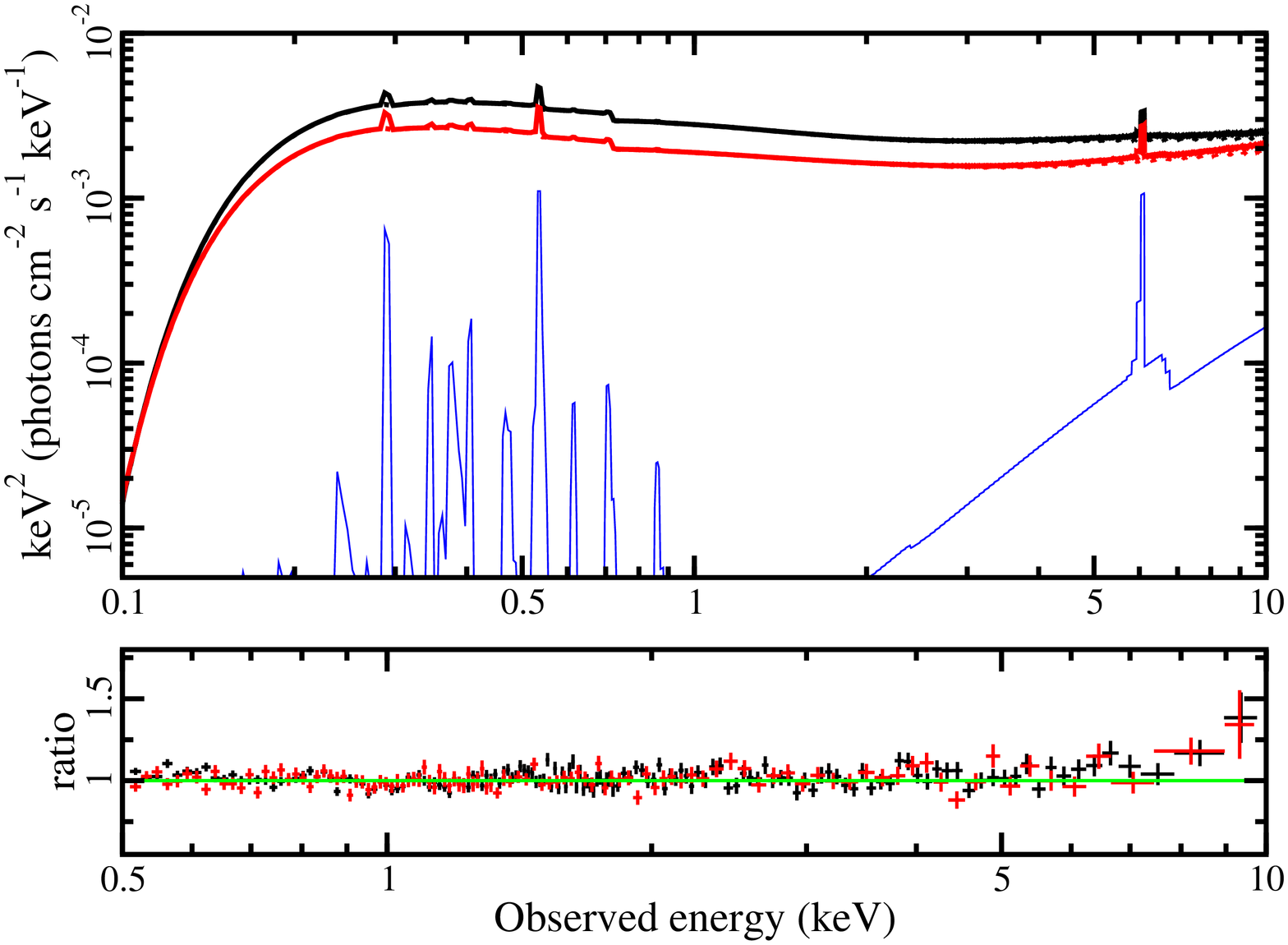}}}
   \caption{The partial covering (top), blurred reflection (middle), and soft Comptonization (bottom) models fit to the flux-reduced spectra with their respective residuals (data/model) shown below each. The high-flux data are black, low-flux are red. The distant reflector, as needed, is shown in blue.}
   \label{FRS_bestabs}
\end{figure}

Allowing only the covering fractions to vary between flux states in the ionized absorption model from Section \ref{iabs} returned \chisq / d.o.f. = 909 / 768 and moderate residual features. There was a small discrepancy in the spectra around 5\keV\ where the model over-predicted the low flux state and under-predicted the high flux state. In addition, only the covering fraction of the denser absorber changed significantly and so the model was revised. With the covering fraction of the denser absorber alone providing the variability, the fit statistic stayed approximately the same for an additional free parameter and the residuals diminished. The model described the data equally well if the column density of the denser absorber was allowed to vary with the covering fraction (Fig.\thinspace\ref{FRS_bestabs}). This scenario makes sense as two overlapping bodies would alter both the total covering fraction and column density along the line of sight. 

A comparative fit was found when the power law parameters alone were allowed free to vary between flux states. It was mentioned above how this scenario is much more difficult to justify: a change in the covering fraction could easily account for short-term variability on kilosecond time scales, but it is challenging to imagine a mechanism that would force the slope of the power law to fluctuate so quickly. Therefore, the best explanation for the short-term variability using absorption is a partial covering model where a cloud of denser material is moving into the line of sight as another cloud with lower density continues to obscure the same solid angle.

A similar procedure was followed for testing the reflection scenario. From past analysis (Section \ref{sec:meanref}) it appears the the X-ray spectra are power law dominated and that changes in the normalization of the primary emitter itself drive most of the short-term variability (Fig.\thinspace\ref{FvarMs}). As the primary emitter brightens or perhaps moves along the axis of rotation, the emissivity profile of the disk can alter (e.g. Wilkins \& Gallo 2015) and material at closer radii experiences a change in ionization. When considering the scenario of a moving ``lamp post" corona emitting isotropically, upward motion would decrease primary X-ray radiation on the inner disk and result in a harder, power law-dominated X-ray spectrum. Downward motion of the corona would cause the disk to be illuminated anisotropically due to light bending, increasing the reflected flux from the inner region and producing a softer, reflection-dominated spectrum. 

The reflection model from Section \ref{sec:meanref} was applied to the flux-resolved spectra (Fig. \ref{FRS_bestabs}). A good fit could be achieved by allowing the power law component to vary in shape and normalization and the blurred reflector to respond with changes in ionization (Table~\ref{BestFRS}).  Allowing other parameters, such as the reflector normalization or emissivity profile, to vary did not enhance the fit.  Most of the change can be attributed to changes in the power law normalization, which seems consistent with the fractional variability and difference spectrum analysis.  Subtle changes in the ionization parameter of the disc arise from changes in the illuminating flux, which can explain the low energy excess in the difference spectrum.

Lastly, the soft Comptonization model (Section \ref{SoftCompt}) was fit to the flux-resolved spectra with only the primary component ($\Gamma$) and the radius of the soft Comptonizing region (\Rcor) allowed to vary. Further \textsc{optxagnf} parameters were thawed gradually until the most reasonable model was found (Table~\ref{BestFRS}). The model that gave the lowest \redchi\* value allowed \Rcor, $kT_{\rm{e}}$, $\tau$, $\Gamma$, and  $f_{\rm{pl}}$ all to vary; however, these parameters values were inconsistent with those from the average spectral fits. The soft Comptonization model that best describes the flux-resolved spectra in a manner consistent with the mean spectral fits allowed only \Rcor, $\Gamma$, and $f_{\rm{pl}}$ to vary. This appears to support the scenario that changes in the primary component are responsible for short-term variability in this object: as the primary component changes over time, the fraction of the primary X-rays that are reprocessed in the soft Comptonizing region ($f_{\rm{pl}}$) also changes, as does the radius of the soft Comptonizing region (\Rcor).

\begin{table*}
\centering
\caption{Summary of models fit to the flux-resolved spectra. Column 1 lists the individual models and Column 2 the model components. Note that only the soft Compton component of the total soft Compton model (Table \ref{BestSC}) is shown for simplicity. The model parameters are shown in Column 3 with their values for the high- and low-flux states listed in columns 4 and 5, respectively. Parameters linked across the epochs are denoted by dots and those without errors are frozen at the given values.}
\scalebox{0.85}{  
		\begin{tabular}{ccccc}
		\hline
		(1) & (2) & (3) & (4) & (5) \\
		 \bf{Model} & \bf{Component} & \bf{Parameters} & \bf{high-flux state} & \bf{low-flux state} \\\hline
		 \\
		Partial Covering & Absorber 1 &$\Gamma$ & $2.40^{+0.04}_{-0.01}$ & \ldots \\
		 Absorption && $N_{\rm{H, 1}}$ (\pscm) & $51.17^{+7.67}_{-8.17}$ & \ldots  \\
		 && $CF_{1}$ & $0.62^{+0.05}_{-0.06}$ & $0.74^{+0.03}_{-0.04}$   \\
		 & Absorber 2 & $N_{\rm{H, 2}}$ (\pscm) & $6.00^{+1.14}_{-1.44}$ & \ldots \\
		 && $CF_{2}$ & $0.32^{+0.06}_{-0.03}$ & \dots \\
		 && \redchi/d.o.f. & 1.18 / 769 \\
		 \\
		 Blurred Reflection & primary power law & $\Gamma$ & 2.14$\pm$0.01 & $2.12\pm0.02$ \\
		 && $F_{0.1-100\keV}$ & 2.17$\pm0.04\times10^{-11}$ & $1.55\pm0.04\times10^{-11}$ \\
		 & blurred reflection & \indexin & 5.67 & \ldots \\
		 && \indexout & 3.0 & \ldots \\
		 && \Rbr (\Rg) & 6.0 & \dots \\
		 && \Rin & 1.79 & \ldots \\
		 && $\theta$\thinspace(\deg) & 43.5 & \ldots \\
		 && $A_{\rm{Fe}}$\thinspace\small{(solar)} & 0.36 & \ldots \\
		 && $\Gamma$ & $\Gamma_{\rm{primary}}$ & \ldots \\
		 && $\xi$\thinspace\small{(\ergcmps)} & $72\pm4$ & $53\pm3$ \\
		 && $F_{0.1-100\keV}$ & 1.12$\pm0.05\times10^{-11}$ & $0.71\pm0.06\times10^{-11}$ \\
		 &reflection fraction & $\mathcal{R}$ & $0.52\pm0.03$  & $0.45\pm0.04$ \\
		 & distant Reflector & $A_{\rm{Fe}}$\thinspace\small{(solar)} & 1.0 & \ldots \\
		 && $\Gamma$ & 1.9 & \ldots \\
		 && $\xi$\thinspace\small{(\ergcmps)} & 1.0 & \ldots \\
		 && \redchi/d.o.f. & 1.12 / 772 \\
		 \\
		 Soft Comptonization & soft Compton & \Mbh\ (\Msun) & 5.9$\times10^{7}$&\ldots \\
		 & component & $D_{\rm{L}}$ (\Mpc) & 233.1 & \dots \\
		 && log(\emph{L} / \Ledd) & -1.05 & \ldots \\
		 && log\Rout & 3 & \dots \\
		 && \emph{a} & 0.8 & \ldots \\
		 && \Rcor\thinspace\small{(\Rg)} & $14.84^{+1.39}_{-1.12}$ & $10.85^{+10.17}_{-1.34}$ \\
		 && $kT_{\rm{e}}$\thinspace\small{(keV)} & $0.61^{+0.20}_{-0.11}$ & \ldots \\
		 && $\tau$ & $7.73^{+1.19}_{-1.27}$ & \ldots \\
		 && $\Gamma$ & 2.12$\pm$0.01 & $1.89^{+0.04}_{-0.03}$ \\
		 && $f_{\rm{pl}}$ & 0.54$\pm$0.02 & $0.44^{+0.05}_{-0.12}$ \\
		 && \redchi / d.o.f. & 1.42 / 772 \\
		 \hline
		\end{tabular}}
\label{BestFRS}
\end{table*}


The spectrum is dominated by a single component and the amplitude and spectral variation are modest in \he0436, thus making more complex timing analysis difficult to justify. That being said, a lag analysis was performed using three different energy bands for comparisons. Given the data quality, lags could only be constrained to within a few hundred seconds in each frequency band. There was no significant detection of a frequency-dependent lag (E. Cackett, private communication), but this could simply be a data quality issue.
 
\section{Discussion} 
\label{Discussion}
The multi-epoch, multi-band analysis of \he0436 has resulted in several scenarios that describe the spectra of the object equally well on a purely statistical basis. In the first interpretation, emission from the central engine passes through at least one ionized body of gas that absorbs the incident radiation. The spectrum produces an apparent soft-excess that is an artifact of the absorption. The second scenario describes the X-ray emission as a composite of multiple sources. Here, the primary source from a compact ``lamp post'' corona dominates over emission from blurred, ionized reflection and a distant, neutral reflector. The \feka\ line seen in this case is a sum of the relativistically broadened iron line originating from the inner regions of the accretion disk and a narrow line from the distant reflector. The soft-excess is a consequence of blurred ionized reflection from the inner regions of the accretion disk. Lastly, it could be that the X-ray spectra of \he0436 is composed of a primary component from a radially extended corona in a ``sandwich'' configuration. The soft-excess in this case is produced by a soft Comptonizing region in the inner disk and a distant reflector produces the \feka\ line.   

It is not unusual for these interpretations to model a spectrum equally well. Due to the complex nature of AGN, it is reasonable to assume that perhaps all three mechanisms (partial covering absorption, blurred reflection, and soft Comptonization) can be present in a single object. However, the degeneracy between the models can lead to confusion when attempting to reconstruct source geometry. Timing analysis can be key in determining the processes that are dominant at a given time. Variability in \he0436 is described well with reflection, as it appears that changes in flux of the primary emitter can be seen influencing both the short- and long-term behavior of this object. Assuming an isotropic lamp-post corona, changes on kilosecond time scales reflect an increase/decrease in primary X-ray flux with respect to the accretion disk. The changes in flux could either be intrinsic or due to the motion of the corona (e.g. Wilkins \et 2014; Gallo \et 2015), which in turn increases/decreases the ionization of the disk at inner radii. Over longer, multi-year time scales fluctuations in power law slope can be seen which would be caused by physical changes in the corona itself (i.e. an alteration in temperature and optical depth). \he0436 remains in a power law-dominated state throughout the observed epochs. When looking at the broader picture of \he0436 and its behavior, the blurred reflection interpretation is the most self-consistent model.

The preference for a reflection scenario in \he0436 leaves the challenge of explaining the persistently low iron abundance. Observations have shown that most AGN have solar to super-solar metallicity values independent of redshift (e.g. Storchi-Bergmann et al. 1998, Hamann et al. 2002). In several independent studies, it was shown that most AGN are also located in massive galaxies with classical bulges (Green \& Ho 2004, Barth et al. 2005). As stellar mass is strongly correlated with metallicity (Tremonti et al. 2004), it is reasonable to expect the majority of AGN to thus have solar metallicity values or higher. The question then arises: under what circumstances would AGN be observed with (apparently) intrinsic sub-solar metallicities?

Groves et al. (2006) set out to find such AGN and searched SDSS observations of Seyfert 2 galaxies to find those that have sub-solar metallicities. Out of a sample of around 23,000 they found only $\sim$40 candidates based on optical line ratios from the NLR. The authors suggest that most AGN are only observed after periods of intense star formation in their host galaxies and so the infalling gas supplying the central engine is already abundant with heavy elements. This implies that AGN with inherently sub-solar metals may somehow be feeding off gas that has yet to be processed by stars. It could be that \he0436 has an unusually low star formation rate for its redshift and/or less Type 1a supernovae to produce iron. Another possibility is that low-metallicity gas from outside of the galaxy has made its way to the central black hole either through a merger or filament, as in the scenarios proposed to explain the unusually low abundances in the NLR of radio-quiet quasar HE 2158-0107 (Husemann et al. 2011).

Other Seyfert 1 galaxies in particular have been known to show sub-solar iron abundances, such as NGC 4593 (Reynolds et al. 2004), NGC 3227 and 4051 (Patrick et al. 2012), and MCG+8-11-11 (Bianchi et al. 2010). Patrick et al. (2012) suggest that a persistently low abundance could represent a component of the \feka\ line originating from a Compton thin material such as the BLR or NLR. The full-width half-maximum (FWHM) of the iron line was calculated for both the pn and merged MOS spectra. The measured FWHM corresponded to velocities of $\sim$0.45\thinspace{$c$} and $\sim$0.31\thinspace{$c$} for pn and MOS, respectively. G10 measured the FWHM of the optical \hb\ line in \he0436 and determined the BLR velocity to be around 3990\kmps or about 0.01\thinspace{$c$}. Thus, while the exact nature of the iron feature in \he0436 remains uncertain, it is clear the line is not originating from the BLR but instead display disk-like velocities ($>$ 0.1\thinspace{$c$}, Miller 2007). A narrow iron feature from a distant reflector is also required to describe the data in this object. The narrow line could originate from the NLR, but it is unlikely that such a significant fraction of the radiation originates from the region (e.g. Kaspi et al. 2002, Fukazawa et al. 2011).

Another alternative is the possibility of spallation reducing the reflected \feka\ flux due to bombardment from high-energy ($>$ 10\thinspace MeV) protons. Skibo (1997) explains how abundances of sub-Fe metals between 4.5 -- 6.4\keV\* can be enhanced in the nuclear regions of AGN at the expense of the Fe line. The author calculates the abundances of metals that could fall in the \feka\* red-wing and finds that the corresponding fluxes (specifically those of Ti, V, Cr, and Mn) relative to Fe would be large enough to raise their lines above the continuum. The combined strength of these sub-Fe lines would be comparable to the strength of the Fe line from which these lines are produced. If this were the case in \he0436, additional features in the red-wing should be detected. In particular, Cr (at 5.4\keV) and Mn (at 5.9\keV) are predicted to have a combined flux $\sim$\thinspace40 - 50\% that of the observed \feka\* line -- no such features are detected in the \xmm spectra.

Finally, perhaps the intrinsic amount of iron in the central engine is not sub-solar, but the amount of other metals higher since the abundance of other metals in the spectral models is fixed at their respective solar values. The iron abundance in this object is low enough, however, that the other metals in the system would have to be super-solar in order to account for the apparent lack of iron. It is a challenge to determine a mechanism that would substantially increase other metals in a gas without affecting the iron, although a lack of Type Ia supernovae (mentioned previously) may do this. More specifically, a decrease in the ratio between Type Ia/Type II supernovae could decrease the relative abundance of iron as both produce the heavy metals that decay into iron, however Type Ia more efficiently disperse material into the surrounding interstellar medium (i.e. no supernova remnant remains).

The possibility of \he0436 having intrinsic sub-solar iron abundance is intriguing as such objects are rare. Deep analysis of the host galaxy would help to further constrain the metallicity of this object and also advance understanding of AGN-host relations. 

\section{Conclusions} 
\label{Conclusions}
A broadband analysis of the Seyfert 1 galaxy \he0436, covering from optical up to $100\keV$ is presented.
SED modelling suggests the source is unobscured and perhaps its black hole mass and distance could be better constrained.  A reverberation mapping campaign would be advantageous to accurately determine the black hole mass in \he0436.

The multi-epoch X-ray spectra are fitted with Comptonisation, blurred reflection, and partial covering models, but do not reveal a significantly preferred model.  Consideration of the rapid variability favors a blurred reflection scenario where the variability is dominated by changes in the power law normalization accompanied by corresponding changes in the disk ionisation.  However, the blurred reflection model indicates a very low iron abundance for \he0436.  The low iron abundance does not seem to be a artefact of the modelling process, and abundances are not tested in the Comptonisation or partial covering scenarios.

Observations of the host galaxy are limited, but would be beneficial to examine any abnormalities in the distribution of abundances in \he0436.  Future X-ray observations with the calorimeter {\em Astro-H} (Takahashi \et 2012) could also reveal narrow features in the X-ray spectra that would allow abundances to be better constrained. In addition to revealing narrow spectral features, \astroh will also allow for simultaneous, broadband soft and hard X-ray observations. This should remove some of the ambiguity that is currently present from utilizing BAT data alongside that from \xmm.


\section*{Acknowledgments}

The authors would like to thank the editors and referee for their valuable feedback that has helped to improve this work. Special thanks to E. Cackett for discussing his lag analysis on this object with us, as well as D. Grupe and C. Done for their time and generous assistance. The authors would also like to thank J. Randhawa for discussing his initial spallation calculations with us. The \xmm\ project is an ESA Science Mission with instruments and contributions directly funded by ESA Member States and the USA (NASA).
\bibliographystyle{mn2e}




\bsp
\label{lastpage}
\end{document}